\documentstyle[floats,multicol,aps,epsf,prb]{revtex} 
\begin{document} 
\draft %%To appear in ``Non Fermi Liquid Physics'', J. Cond. Matt. (1997). 
% ***********    This is for two columns ******************************* 
\twocolumn[\hsize\textwidth\columnwidth\hsize\csname @twocolumnfalse\endcsname 
\title 
{``How should we interpret the {\sl two} transport relaxation times 
in the cuprates ?''} 
\author{P. Coleman$^1$, A. J. Schofield$^1$ and A. M. Tsvelik$^2$} 
\address{ 
$^1$Serin Laboratory, Rutgers University, P.O. Box 849, 
Piscataway, New Jersey 08855-0849} 
\address{ 
$^2$Department of  Physics, University of Oxford, 1 Keble Road, 
Oxford OX1 3NP,  UK} 
\maketitle 
\date{\today} 
\maketitle 
%\widetext 
\begin{abstract} 
We observe that the appearance of two transport relaxation times in 
the various transport coefficients of cuprate metals may be understood in 
terms of scattering processes that discriminate between currents that 
are even, or odd under the charge conjugation operator.  We develop a 
transport equation that illustrates these ideas and discuss its 
experimental and theoretical consequences. 
\end{abstract} 
 
\vskip 0.2 truein 
\vskip2pc] 
 
\vskip 0.5truein 
{\sl `` There are and can exist but two ways of investigating and discovering 
truth.  The one hurries on rapidly from the senses and particulars 
to the most general axioms, and from them, as principles and their 
supposed indisputable truth, derives and discovers the intermediate 
axioms.  This way is now in fashion. 
 
The other constructs its axioms  from the senses and  particulars, 
by ascending continually and gradually, till it finally arrives 
at the most general axioms.  This is the true, but as yet 
 untried way.'' 
\vskip 0.2truein 
 
\hskip 0.5 truein Francis Bacon in ``Novum Organum'', 1620\hfill } 
\vskip 0.5truein

\section{{\bf Introduction.}} 
 
%%\begin{enumerate} 
%% 
%%\item  Phenomenological: a Product of relaxation times, very difficult to incorporate within 
%%    a conventional metallic framework. 
%% 
%%\item 
%%The basic controversy: is two-time interpretation 
%%reasonable? T^2 behavior a true scattering rate, 
%% or is it derived from a complex interplay of other scattering mechanisms. 
%% i.e  Does the Hall current really decay at T^2 ? 
%% 
%%For example, skew scattering.....   1/T and T 
%%                Fermi surface....   T and $T^{1.5}$ 
%%Both have difficulties. 
%% 
%%\item  Hall sum rule.   Very difficult to escape the conclusion that the 
%%d. Hall angle determines the Hall decay. 
%% 
%%\item  But if the interpretation is good, how are we to understand it? 
%% 
%% 
%% 
%%\item 
%%What symmetry constraints are imposed on the scattering mechanism at 
%%the the Fermi surface (if one exists) 
%%if indeed the two-time interpretation is correct? 
%% 
%%\item 
%% Discussion of scattering around Fermi surface:  one photon- fast decay 
%%    two photons- slow  decay.  How? 
%% 
%%\item 
%% Another way to see its remarkable: discrimination between Lorentz electric 
%%    field E= vxB and Electric field. 
%% 
%%\item   Discrete symmetries of the photon. 
%%  Symmetries of the Fermi surface: condensed matter physics version of 
%%    discrete relativistic symmetries.  Photon is odd under these. 
%% 
%%\item 
%% Of the three symmetries, only one, C, discriminates between j-hall and j-tr 
%%(Furry's theorem) 
%% 
%%\end{enumerate} 
 
One of the striking features of the cuprate metals, is the 
appearance of two, qualitatively different transport relaxation 
times.    Resistivity and optical measurements indicate that 
electric currents relax at a rate which grows linearly\cite{linear} with 
temperature \begin{eqnarray} 
\Gamma_{tr} \sim \eta T, \qquad (\eta \sim 2). 
\end{eqnarray} 
By contrast, 
the ``Hall relaxation rate'' obtained from the  Hall angle\cite{chien} 
$ 
\theta_H = {\omega_c/\Gamma_H}, 
$ where $\omega_c  = e H/m $ 
is the cyclotron frequency, 
shows a qualitatively different quadratic temperature dependence 
\begin{eqnarray} 
\Gamma_H = {T^2 \over W} + n_i b\; . 
\end{eqnarray} 
The quadratic form of $\Gamma_H$ is robust against a 
finite concentration of impurities $n_i$. 
Estimates of $W$ 
based on d.c. measurements by Ong et al., give $W\sim1000K$, but more 
recent direct measurements of $\tau_H$ from the A.C. Hall angle, suggest that 
it may be significantly smaller. 
These features led Anderson, some five years ago, to conjecture that there are 
two  transport relaxation times in the cuprate metals which independently 
govern the decay of electrical and Hall currents. \cite{phil,romero} Although 
subsequent experimental results
have tended to reinforce this phenomenological 
interpretation, Anderson's proposal remains highly controversial. 
 
This article discusses the  two-relaxation time conjecture.  We 
sharpen the definition of the Hall and electric transport relaxation rate 
and show that a sum rule for the Hall angle means that 
Anderson's interpretation can be made without 
reference to the microscopic physics.    We explain why the 
robustness of 
the quadratic temperature dependence to changes in hole and impurity 
concentrations makes it very difficult to embrace the 
various alternative interpretations of the magneto-transport: 
the quadratic temperature dependence of $\Gamma_H$ 
appears there 
as a fortuitous cancelation of independent scattering processes. 
 
Motivated by these considerations, 
we then  pursue the consequences of the Anderson conjecture, 
bringing symmetry considerations into play.\cite{ourprl}  We  identify 
charge conjugation as the key symmetry 
distinction between the electric and the Hall currents, and argue 
that in the cuprates, there must be  scattering processes  involving 
the emission of charge that cause degenerate electron and 
hole states to admix in the normal state. 
These  ideas are illustrated by 
a phenomenological transport equation. In the final section, 
we discuss the challenge of isolating the  physics that 
simultaneously accounts 
for both the marginal scattering of the electrons and the 
delineation of electric and Hall currents. 
 
\section{Review of Experimental results} 
 
The inverse-square  Hall angle $\theta_H\sim T^{-2}$ 
is a robust feature of the cuprate 
metals which governs both the Hall conductivity and 
the magneto-resistance.  Remarkably, 
the two relaxation times enter {\sl multiplicatively} 
into the transport coefficients. 
The Hall conductivity $\sigma_{xy}=\sigma_{xx}\theta_H$  has the form 
\begin{eqnarray} 
\sigma_{xy} \propto   {H \over \Gamma_{tr} \Gamma_{H}},\qquad\qquad &(\sim { T^{-3}}) 
\label{look} 
\end{eqnarray} 
where $H$ is the external field strength. 
In optimally doped cuprates, the 
magneto-conductivity $\Delta \sigma_{xx}$ depends 
quadratically on the Hall angle\cite{ong} 
\begin{eqnarray} 
\Delta \sigma_{xx} \propto   {H^2 \over \Gamma_{tr} \Gamma_{H}^2},\label{look2} 
\qquad\qquad &(\sim T^{-5}) 
\end{eqnarray} 
In a normal metal  there is {\em one} transport 
relaxation rate $\Gamma_{tr}(\vec p)$ at each point in momentum space. 
$\Gamma_{tr}(\vec p)$  can have strong momentum dependence, 
but since transport is a zero-momentum probe, 
momentum conservation {\sl prevents a multiplicative 
combination of 
scattering rates from different 
points} on the Fermi surface. 
For example,  the  Hall conductivity  of a Fermi liquid 
is given by the second moment of the transport relaxation time, 
around the Fermi surface 
\begin{eqnarray} 
\sigma_{xy} \propto \int dp_z \int_{\rm FS} 
{\vec v \times d \vec v \over 
(\Gamma_{tr}(p))^2 } \; , 
\end{eqnarray} 
where FS denotes a line integral around the Fermi surface 
in the plane perpendicular to the field, $\vec v$ is the 
Fermi velocity and $d \vec v = d\vec p \cdot \nabla \vec v$ 
is the change in $\vec v$ along the line.\cite{onghall} 
Momentum conservation obliges us to interpret 
a multiplicative 
combination of $\Gamma_{tr}$ and $\Gamma_H$ in terms 
of two relaxation times at the same point in  momentum space. 
 
A clear manifestation of these 
two relaxation times, is the violation of 
Kohler's rule. 
In conventional metals, where 
$\Gamma_H\sim \Gamma_{tr}$, the transverse magneto-resistance 
obeys Kohler's rule $\delta \rho/\rho \propto (H/\rho)^2$.\cite{ziman} 
Kohler's rule is violated in the optimally doped and 
under-doped cuprate metals.\cite{ong} Instead, 
the appearance of the same Hall angle in  the 
magneto-resistance means that a modified rule 
$\delta \rho/\rho \propto (H/\Gamma_H)^2\propto (HR_H/\rho)^2$ 
is approximately satisfied.\cite{tanake} 
Recent studies show that when the cuprate metals are over-doped, 
Kohler's rule behavior is restored.\cite{takagi} 
 
\begin{figure}[tb] 
% ********   This is for two columns 
\epsfxsize=3.5in 
% ***********For one column  ******************** 
%\epsfysize=5.5in 
% ***********************************8 
\centerline{\epsfbox{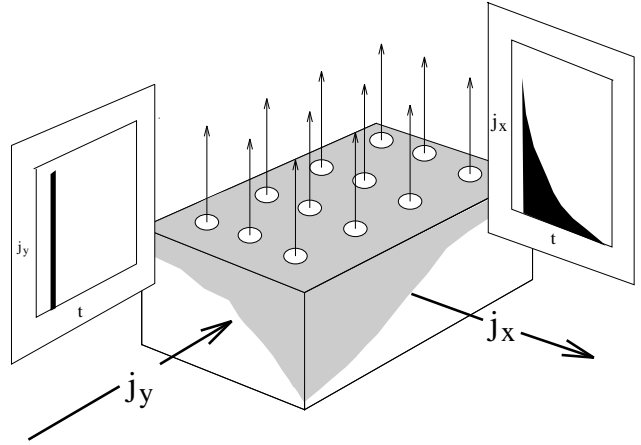}} 
\vskip 0.3truein 
\protect\caption{ 
Illustrating the Hall response $j_x(t)= j_o\Theta_H(t)$ 
to an input current pulse $j_y(t) = j_o\delta(t)$. } 
\label{Fig1} 
\end{figure}

To pursue this discussion, we need a clear 
idea of what we mean by  the  Hall and electric current relaxation rates. 
Electric current is the 
response to 
an applied electric field, 
\begin{eqnarray} 
j_x(t) = 
\int_{-\infty}^{t} \sigma_{xx}(t-t') E_x(t') dt' \; , 
\end{eqnarray} 
where $\sigma_{xx}(t-t')$ is the Fourier transform of the 
frequency dependent conductivity. 
When we speak of a relaxation rate for the electrical current, we 
mean that the current response function  $\sigma_{xx}(t-t')$ 
has an exponential form 
\begin{eqnarray} 
\sigma_{xx}(t-t') = {ne^2 \over m} e^{-\Gamma_{tr}(t-t')} \; , 
\end{eqnarray} 
This is the origin of the Drude peak in the optical conductivity. 
Hall current is the retarded transverse response to 
to an input  current (Fig.~\ref{Fig1} ) 
as follows 
\begin{eqnarray} 
j_{x}(t)= 
\int_{-\infty}^{t} \Theta_H(t-t') j_y(t') dt' \; , 
\end{eqnarray} 
where $\Theta_H(t)$ is the Fourier transform of the 
frequency dependent Hall angle $\theta_H(\omega)= 
\sigma_{xy}(\omega)/\sigma_{xx}(\omega)$. 
The Hall relaxation rate refers to the decay of the Hall 
current in response to a sudden pulse of current 
\begin{eqnarray} 
\Theta_H(t-t') = \omega_c e^{-\Gamma_H (t-t')} \; , \qquad (t>t') \; . 
\end{eqnarray} 
These are the operational definitions of $\Gamma_{tr}$ and 
$\Gamma_H$. 
 
Recent experimental advances  make it possible 
to  directly  probe  this Hall decay rate using 
optical transmission  experiments. 
Such measurements by 
Kaplan et al\cite{drew} show 
that the frequency dependent 
Hall angle can be fit to a single Lorentzian form: 
\begin{eqnarray} 
\theta_H(\omega) 
  = \left[ {\omega_c  \over \Gamma_H - i\omega}\right] \; , 
\end{eqnarray} 
where $\Gamma_H$ is smaller than the transport relaxation rate. 
Measurements on $YBCO$ films at 100K indicate that 
$\Gamma_H\sim {1 \over 4} \Gamma_{tr}$ at this temperature.  At 
present, no detailed 
measurements of the temperature 
dependence of $\Gamma_H$ are available. 
 
In this connection  however, 
there is an important optical sum rule 
\cite{drewcoleman} 
\begin{eqnarray} 
2 \int_0^{\infty} {d \omega \over \pi} \theta_H(\omega) = \omega_c \; . 
\end{eqnarray} 
This is the transverse counterpart of the forward 
or ``f-sum'' rule. 
The transverse, or ``t-sum rule'' 
 is exact, if all frequencies are included. 
For a single, split off-band, there is an effective sum rule 
with a cyclotron frequency 
\begin{eqnarray} 
\omega_c = 
2eH { \sum {\rm det} ( \underline{\rm m}_p^{-1} )n_p 
\over\sum {\rm Tr}  ( \underline{\rm m}_p^{-1} ) n_p} \; , 
\end{eqnarray} 
which is determined by a sum over the entire band.  This 
quantity is expected to be almost temperature independent. 
Since $\theta_H^{dc} \propto 1/T^2$,  the width of the hall spectral function 
$\theta_H(\omega)$ {\sl must } grow quadratically with temperature 
to preserve the sum rule. 
Measurements reported at this meeting\cite{newdrew} seem to provide 
some of the first direct   support for this conclusion. 
It should be clear from this discussion, that 
the two-relaxation time interpretation of the cuprate 
transport can be made purely on the basis of the 
sum rule, and experimental observation. 
 
Anderson's interpretation of the Hall mobility in terms of 
two relaxation times  nevertheless poses a serious paradox for 
the microscopic physics. 
The problem is, that in a conventional metal 
there  is no {\sl microscopic} distinction between a ``Hall'' 
and an ``electric'' current.  Electrons 
respond to the total Lorentz electric field 
\begin{eqnarray} 
\vec {\cal E} = \vec E + \vec v \times \vec H \; , 
\end{eqnarray} 
where $\vec v$ is their group velocity. Electric and magnetic 
fields always enter the transport equations in this combination, 
so that electrons on the Fermi 
surface can not tell external electric and internal Lorentz forces 
apart. 
Anderson has suggested that one way to produce two such autonomous 
scattering rates is to develop spin-charge decoupling.  In his 
picture, the $T^2$ scattering rate is associated with spin excitations, 
or ``spinons'', whilst the linear relaxation rate is associated with 
charge excitations, or ``holon''.  Anderson suggests that 
the motion of spinons 
produces a charge back-flow which is responsible for the Hall current. 
What is lacking is an explanation of 
why this charge back-flow only carries a Hall current. 
To pursue the idea of two transport relaxation times 
we need to find 
a symmetry reason for this selectivity.

Before taking this path 
it is instructive to consider the alternatives. 
Two classes of 
proposal have been made: 
\begin{itemize} 
 
\item Strong momentum-dependent scattering.\cite{carrington,pines,hlubina} 
 
\item Skew scattering.\cite{kotliar} 
 
\end{itemize} 
The first scenario\cite{carrington,pines,hlubina} envisions two 
distinct regions of the Fermi surface: a ``hot spot" 
where the scattering rate is a linear function of temperature, 
$\Gamma_{{\rm hot}}\propto T$, and a second ``{{\rm cold}}" region of 
the Fermi surface with a weaker temperature dependence of the 
transport relaxation rate $\Gamma_{{\rm cold}}$. 
This theory presumes that the ``hot spot" 
dominates the electrical conductivity, 
whereas the {\rm cold} region, with a higher Fermi surface curvature, 
sets  the Hall conductivity, so that 
\begin{eqnarray} 
\sigma_{xx}&\propto& {1 \over \Gamma_{{\rm hot}}} \; , \cr 
\sigma_{xy}&\propto & {H \over (\Gamma_{{\rm cold}})^2 } \; . 
\end{eqnarray} 
In this interpretation, the quadratic temperature 
dependence of the Hall relaxation rate is not fundamental, 
but appears as  a 
consequence of a cancelation between relaxation rates at different 
parts of the Fermi surface: 
\begin{eqnarray} 
\Gamma_H \propto{\Gamma_{{\rm cold}}^2 \over 
\Gamma_{{\rm hot}}}\; ,\qquad (\hbox{\rm ``hot \ spot \ scenario''} ). 
\end{eqnarray} 
If $\Gamma_{{\rm cold}}\propto T^{1.5}$, then $\Gamma_H\propto T^2$. 
 
The ``skew scattering" interpretation of the anomalous 
Hall angle, presumes the presence of chiral current 
fluctuations which, through their coupling to the 
electrons cause a field-dependent ``skew scattering" component 
to develop in the inelastic scattering.  This component 
is required to have a singular dependence on temperature of the 
following form 
\begin{eqnarray} 
\Gamma_{\rm skew} \propto {H \over T} \; . 
\end{eqnarray} 
The skew-scattering renormalizes the effective cyclotron frequency 
without changing the current relaxation rates, changing 
$\omega_c \rightarrow \omega_c^* = \omega_c + \Gamma_{\rm skew}$, 
so that for an almost compensated band 
 $\omega_c^* \sim \Gamma_{\rm skew}$ giving 
\begin{eqnarray} 
\Gamma_H(T)\propto {\Gamma_{\rm tr} \over \Gamma_{\rm skew}} \; , 
\qquad ({\rm skew\ scattering}) 
\end{eqnarray} 
is a quadratic function of temperature. 
 
These alternate scenarios face a serious common difficulty. 
Each case 
requires the presence of a  fortuitous cancelation of two 
independent scattering processes.  The temperature dependences 
$\Gamma_{\rm hot} \propto T^{1.5}$ and $\Gamma_{\rm skew} \propto 1/T$ 
are actually more complicated than the simple quadratic 
relaxation rate they are supposed to give rise to. 
Furthermore in these 
theories, if  the linear transport relaxation rate 
is substantially modified by the addition of impurities 
or changes in the hole concentration, then the quadratic 
temperature dependence of $\Gamma_H$ is lost.  This is 
not what is seen: changes in impurity concentration or 
oxygen doping which eliminate the linear temperature 
dependence of the resistivity do not change the 
the quadratic temperature dependence of 
$\Gamma_H$. These features all tend to suggest that $\Gamma_H$ is 
a truly  autonomous scattering rate, not a fortuitous cancelation 
of other scattering processes. 
 
So to conclude this section, rather general considerations of an experimental 
and theoretical nature appear to force us to 
return to Anderson's original conjecture and to ask 
what general 
constraints it places on the microscopic physics.  This is the subject 
of the remainder of this paper.

\section{\bf Symmetries of the Fermi surface} 
 
What type of scattering event can lead to different Hall and electric 
relaxation rates?  Fundamentally, electrical and Hall currents differ 
in the number of photons absorbed. 
In linear response theory, electric current 
involves the absorption of a single photon with a finite frequency.  The 
Hall current is generated by a two-photon absorption process: the 
first photon excites an electron-hole pair about the Fermi surface, the 
second transfers momentum, causing the electron and hole 
in the pair to precess around the Fermi  surface. 
The diagrammatic expression of the electric and Hall conductivities involves 
the bubble and triangle diagrams shown in 
Fig.~\ref{Fig2}\footnote{For a discussion of the calculation of the 
Hall conductivity see Appendix A and reference~\onlinecite{voruganti}.} 
\begin{figure}[tb]. 
% ********   This is for two columns 
\epsfxsize=3.4in 
% ***********For one column  ******************** 
%\epsfysize=5.5in 
% ***********************************8 
\centerline{\epsfbox{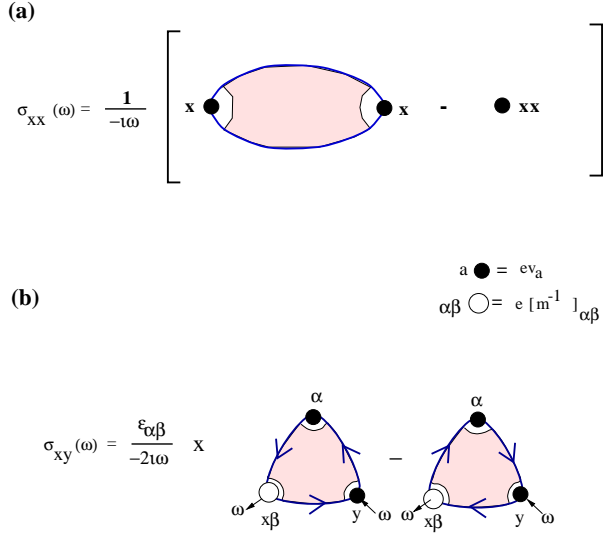}} 
\vskip 0.3truein 
\protect\caption{(a) ``Bubble diagram entering conductivity, 
(b) ``Triangle diagrams'' entering into the Hall conductivity. 
Open circle denotes the effective mass tensor which appears 
when the derivative of the velocity operator with respect 
to the external momentum is taken, to extract 
the magnetic field dependence (See Appendix A).   Filled circle 
denotes the velocity operator. 
} 
\label{Fig2} 
\end{figure} 
Heuristically 
\begin{eqnarray} 
\sigma_{xx}(\omega) &\sim & {{\rm bubble\ diagram} 
(\omega)\over -i \omega},\cr 
\sigma_{xy}(\omega) &\sim &{1\over  \omega} 
\bigl[ 
{\rm triangle \ diagram} 
(\omega)\bigr] . 
\end{eqnarray} 
Each external leg 
of the diagram carries the same momentum, and thus the 
relaxation times which enter multiplicatively into the 
Hall conductivity must be derived from the {\sl same point} 
in momentum space. 
How then can a two-photon process involve the product of two 
qualitatively different relaxation time-scales? 
 
The shrewd diagrammatician will recognize that, in general, we have 
to include 
vertex corrections and that 
if these are important a simple relaxation time discussion 
of the problem may not be valid. Recall however, that 
the qualitative simplicity of the experiment suggests that the 
underlying physics {\sl does} have some kind of 
simple relaxation time interpretation, albeit not the 
one we are familiar with.  In conventional metals 
the effect of vertex corrections is to replace the 
electron inelastic scattering rate by the appropriate transport 
relaxation rate 
\def\joinrel{\mathrel{\mkern-8mu}} 
\def\joinreld{\mathrel{\mkern-20mu}} 
\def\joinreldd{\mathrel{\mkern-40mu}} 
\def\longarrow{ 
\joinrel\relbar 
\joinrel\relbar 
\joinrel\relbar 
\joinrel\relbar 
\joinrel\relbar 
\joinrel\relbar 
\joinrel\relbar 
\joinrel\relbar\joinrel\longrightarrow} 
\begin{eqnarray}\Gamma(\omega, \vec p) 
\ \ \mathrel 
{\mathop{\longarrow}^{\rm vertex \ eff. }}\ 
\Gamma_{\rm tr}(\omega, \vec p) \; , 
\qquad (\rm simple \ metal) \; . 
\end{eqnarray} 
Experiments motivate us to seek 
an explanation where the effect of the vertices is 
to produce two relaxation times on the external legs 
\begin{eqnarray}\Gamma(\omega, \vec p) 
\ \ \mathrel 
{\mathop{\longarrow}^{\rm vertex \ eff. }}\ 
\left\{ 
\begin{array}{rl} 
\displaystyle\Gamma_{\rm tr}&\displaystyle(\omega, \vec p) \; ,\cr 
\displaystyle\Gamma_{\rm H}&\displaystyle(\omega, \vec p) \; . 
\end{array} 
\right.\label{tryit} 
\end{eqnarray} 
We seek an explanation where the one 
photon absorption process involves the fast relaxation rate $\Gamma_{\rm tr}$, 
but the two-photon absorption process, involves a product 
of both relaxation times. 
 
There is a vital distinction between the electrical and Hall current. 
Magnetic fields couple to the 
the  momentum dependent part of the 
current operator, and it is this component 
that gives rise to a Hall response. 
Operationally, the momentum derivative which enters 
in the Hall conductivity acts on the vertices of the 
Hall conductivity, 
extracting the momentum dependent part of 
the current operator. 
If we are to understand 
the two-relaxation time interpretation then we must 
understand what feature of the scattering processes 
can force the momentum 
dependent part of the current operator to develop a different 
relaxation rate. 
 
We now  examine the 
symmetries that  delineate  Hall and electric current. 
There are  three fundamental symmetry operations 
that describe the excitations around a Fermi surface: 
time-reversal  ($T$), inversion ($P$) 
and charge conjugation symmetry ($C$). 
Consider a Fermi surface where the kinetic energy is 
defined by the Hamiltonian 
\def\rp{\rm \vec p} 
\begin{eqnarray} 
H_o= \sum_{\rm p \sigma} \epsilon_{\rp } \psi{^{\dag}}_{\rp \sigma} 
 \psi_{\rp \sigma} \; . 
\end{eqnarray} 
To make our discussion precise, we 
suppose that throughout any calculation of current response 
functions, we may approximate the 
Fermi surface by a polyhedron where the center of each face is the Fermi 
wave-vector $\vec p_F$. At the end of the calculation, the number 
of sides of the polyhedron is to be taken to infinity. 
Consider an electron state with momentum $\vec 
p = \vec p_F+\delta \vec p$, near a Fermi surface.  This state is 
degenerate with  a hole state formed by annihilating an electron at 
momentum $\vec p^*= \vec p_F - \delta \vec p^*$  ($\epsilon_{\rm \vec p} 
= - \epsilon_{\rm \vec p^*}$) . The operation that links 
the two states is that of charge conjugation. 
It is also degenerate with the up and down 
electron states at momentum $-\vec p$ that we obtain by 
space or time reversal operations. 
We define the set of three symmetry 
operations as follows 
\footnote{ Note that the labels we have assigned to these 
symmetry operations are specific for a Fermi surface, and do not 
precisely corresponds to the $P$, $C$ and $T$ operators of 
relativistic quantum field theory.} 
: 
\begin{eqnarray} C{^{\dag}} \psi_{\vec p \sigma}C &=& \sigma 
\psi{^{\dag}}_{\vec p^* \ -\sigma},\cr P{^{\dag}} \psi_{\vec p \sigma}P 
&=& \psi_{-\vec 
p \sigma},\cr T{^{\dag}} \psi_{\vec p \sigma}T &=& 
\sigma \psi_{-\vec p\ -\sigma}. 
\label{cpt} 
\end{eqnarray} 
\begin{figure}[tb] 
% ********   This is for two columns 
\epsfxsize=3.4in 
% ***********For one column  ******************** 
%\epsfysize=5.5in 
% ***********************************8 
\centerline{\epsfbox{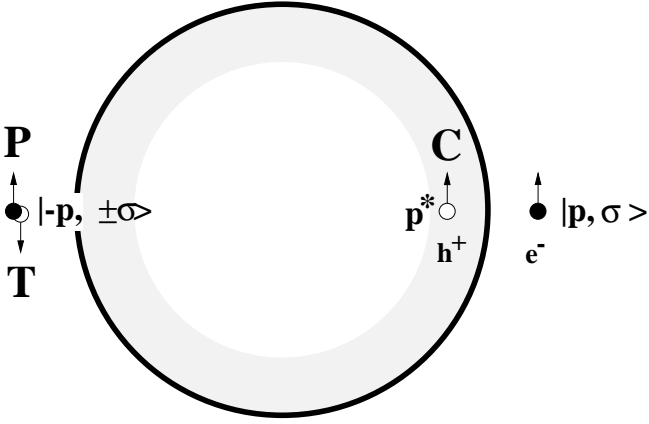}} 
\vskip 0.3truein 
\protect\caption{Illustrating the action of operators $C$, $P$ and $T$ 
on the electron state $|\vec p, \sigma\rangle$. 
 } 
\label{Fig3h} 
\end{figure} 
Charge conjugation changes the charge of the quasiparticle, 
but does not alter it's velocity.  Parity and time reversal 
change the velocity of the quasiparticle, (Fig.~\ref{Fig3h}) without changing 
its charge, as follows: 
\begin{eqnarray} 
\hat C: \qquad {\rm e} &\longrightarrow& -{\rm e},\cr 
\hat P, \ \hat T: \qquad {\vec v} &\longrightarrow& -{\vec v}. 
\end{eqnarray} 
Time-reversal also flips the spin of the quasiparticle. 
These operators   span a manifold of eight degenerate 
electron and hole states on opposite sides of the 
Fermi surface. Provided the relevant physics 
involves electron states near enough to 
the Fermi surface, we expect these symmetries to be  preserved 
by interactions. This is the case, for instance, in the one-dimensional 
Luttinger liquid. These symmetries are also preserved in the 
presence of a magnetic field, where 
\begin{eqnarray} 
H_o[\vec A]= \sum_{\rm p \sigma} \epsilon_{\rp - e\vec A} \psi{^{\dag}}_{\rp 
\sigma} 
 \psi_{\rp \sigma}\; . 
\end{eqnarray} 
Using the properties $\epsilon_{\rp} = \epsilon_{-\rp} = 
-\epsilon_{-\rp ^*}$, the 
Hamiltonian transforms as follows 
\begin{eqnarray} 
{\cal O}{^{\dag}} H[\vec A]{\cal O} =H[ -\vec A],\qquad ({\cal O} = C, \ P, 
\ T), 
\end{eqnarray} 
so that these transformations are equivalent to changing 
the external field 
$\vec A \longrightarrow \vec A^*= -\vec A$, or more correctly 
\begin{eqnarray} 
\vec A^*(x,t)=\left\{ 
\begin{array}{rcl} 
\displaystyle 
 -\vec A(x,t),&\qquad\qquad&(C)\cr 
-\vec A(-x,t),&\qquad\qquad&(P)\cr 
-\vec A(x,-t).&\qquad\qquad&(T) 
\end{array}\right. 
\end{eqnarray} 
Suppose   $\vert \Psi_o\rangle$ is a current-free state 
of the normal metal.  If an external field is applied 
to this state, the current carrying state which develops is 
\begin{eqnarray} 
\vert E, H \rangle ={\cal  T} e^{i \int (\vec A \cdot \vec j) 
d^3xdt} 
\vert \Psi_o\rangle, 
\end{eqnarray} 
where ${\cal T}$ denotes the time-ordered product, 
$\vec E$ and $\vec H$ are the electric and magnetic 
fields. Now consider the transformed   state 
\begin{eqnarray} 
\vert E^*, H^* \rangle= {\cal O}\vert E, H \rangle. 
\end{eqnarray} 
Since $\vert \Psi_o\rangle$ carries no current, we expect 
it to be symmetric under $P$, $C$ and $T$, 
${\cal O}\vert \Psi_o\rangle= \vert \Psi_o\rangle$. 
Thus 
\begin{eqnarray} 
\vert E^*, H^* \rangle 
={\cal  T} e^{i \int (\vec A^* \cdot \vec j )d^3xdt} 
\vert \Psi_o \rangle \; , 
\end{eqnarray} 
corresponds to the state which would evolve in response 
to the  transformed electric and 
magnetic fields, 
$\vec E^*  = - \partial \vec A^*/ \partial t$ 
and  $\vec H^* = \vec \nabla \times \vec A^*$.  The 
parities of these fields  under $C$, $P$ and $T$ are then 
\begin{center} 
\begin{tabular}{|c|c|c|c|} 
\tableline 
$\quad$&$\quad A \quad$&$\quad E\quad$& 
$\quad H\quad$\\ \tableline 
C&-&-&-\\ 
P&-&-&+\\ 
T&-&+&-\\ 
\tableline 
\end{tabular} . 
\end{center} 
We can work out the parities of the 
electric $\vec {\cal J}_E$ and Hall current 
$\vec {\cal J}_H$ operators under these various transformations 
by comparing their expectation values in the 
two states $\vert E, H \rangle$ and $\vert E^*, H^* \rangle$, 
\begin{eqnarray} 
{j}^*_{E,H} &=& \langle E^*, H^*\vert 
{\cal J}_{E,H}\vert E^*, H^*\rangle \; ,\cr 
j_{E,H}&=& \langle E, H\vert 
{\cal J}_{E,H}\vert E, H\rangle. 
\end{eqnarray} 
To find the parities, 
we may apply the transformations 
$(E,  \ H) \rightarrow (E^*,  \ H^*) 
$ 
to the classical equations of motion, 
\begin{eqnarray} 
\vec {j}_E &\sim&{ne^2 \over m} \int \vec E dt,\cr 
\vec {j}_H &\sim &{e \over m}\int \vec {j}_E \times \vec H  dt. 
\end{eqnarray} 
It follows that  electric 
current transforms in the same way as the 
vector potential $\vec {\cal J}_E\sim \vec A$, whereas 
Hall current transforms in the same way as the 
Poynting vector $\vec E\times \vec H$. 
The parities of these two currents are thus given by the following 
table 
\begin{center} 
\begin{tabular}{|c|c|c|} 
\tableline 
$\quad$&$\quad {\cal J}_E\sim \vec A \quad$& 
$\quad 
{\cal J}_H\sim  \vec E \times\vec H\quad$\\ \tableline 
C&-&+\\ 
P&-&-\\ 
T&-&-\\ 
\tableline 
\end{tabular} .
\end{center} 
Charge conjugation parity {\sl  uniquely} discriminates between 
Hall and electric currents. 
 
We may confirm the results of this 
heuristic discussion by directly applying the transformation operators 
to the total current operator.  Under $\hat P $ and $\hat T$, 
the entire current operator transforms in the same way, and 
it is only under $\hat C$ that it divides up into two components 
of opposite parity. 
 
Using this information we can 
construct the Hall and electric current operators as follows. 
Suppose the wave-function for a metal in an electro-magnetic field 
is $\vert\vec E, \vec H \rangle$. We may construct the 
state in which the electric and magnetic fields are 
reversed by applying the charge conjugation operator 
\begin{eqnarray} 
\vert-&\vec E,\  -\vec H \rangle 
&= \hat C \vert\vec E, \vec H \rangle. 
\end{eqnarray} 
The current in this state is given by 
\begin{eqnarray} 
\langle -\vec E,\ -\vec H\vert 
\vec {\cal J} \vert-\vec E,\  -\vec H \rangle \cr 
= \langle \vec E,\ \vec H\vert 
C{^{\dag}}\vec {\cal J}C \vert\vec E,\  \vec H \rangle. . 
\end{eqnarray} 
In the original state $\vert E, H \rangle$ state,  the total
current is given 
by a sum of electric and Hall currents 
\begin{eqnarray} 
\langle \vec E,\ \vec H\vert 
\vec {\cal J} \vert\vec E,\  \vec H \rangle = \vec j_E +\vec 
j_H. 
\end{eqnarray} 
In the state with reversed fields, 
the electric current reverses, but the Hall current does not, i.e. 
\begin{eqnarray} 
\langle \vec E,\ \vec H\vert 
C{^{\dag}}\vec {\cal J}C \vert\vec E,\  \vec H \rangle = 
-\vec j_E +\vec 
j_H. 
\end{eqnarray} 
It follows that the uniform electric and Hall current operators may be 
defined as follows 
\begin{eqnarray} 
 \vec {\cal J} _ { \textstyle \left\{ 
{E \above 0pt H } 
 \right\}} 
 &=& {1 \over 2} [\vec {\cal J} \mp C{^{\dag}}\vec {\cal J}C ]. 
\end{eqnarray} 
When we  explicitly evaluate these operators near the Fermi 
surface, we find that 
\begin{eqnarray} 
\vec {\cal J} _E & =&  e \sum_{\rp \sigma} 
\vec v_F \psi{^{\dag}}_{\rp  \ \sigma} \psi_{\rp \ 
\sigma}, \qquad(C=-1)\cr 
\vec {\cal J} _H  & =&  e \sum_{\rp \sigma} 
 \psi{^{\dag}}_{\rp  \ \sigma} 
\underline{\rm m}^{-1}{(\delta \rp - e \vec A)} 
\psi_{\rp \sigma}, \qquad(C=+1) 
\end{eqnarray} 
where we have expanded the electron  velocity in terms of the 
effective mass tensor near the Fermi surface. The 
Hall current is zero when the Fermi surface is flat. 
 
Notice how 
the electric and Hall current 
depend on changes in the electron occupation that are respectively 
even and odd about the Fermi surface.   Magnetic photons 
introduce a small shift in the momentum $\delta q$ 
and this gives rise to Hall current proportional to $\delta 
q$. 
If Hall and electric currents 
relax at qualitatively different rates then there must be 
scattering processes which selectively relax these two different types 
of quasiparticle distribution. 
In other-words,  the scattering, and hence the microscopic 
self-energies of the electrons must depend on  charge 
conjugation symmetry.  Schematically, if $\underline {\Sigma}$ 
is the electron self energy, then we require 
\begin{eqnarray} 
\underline {\Sigma}= {\Sigma}_1 + {\Sigma}_2 \tilde C,\label{goit} 
\end{eqnarray} 
where $\tilde C$ is one of $C$, $CP$, $CT$ or $CPT$. 
 
There are two points to discuss about this conclusion. 
First, we may reject the possibilities $\tilde C= CP$ or 
$\tilde C = CT$, because under these transformations 
the electric current is an even-parity operator and the 
Hall current is an odd-parity operator. 
We shall shortly see that even-parity operators are always 
``short-circuited'' by the quasiparticles with the 
slowest relaxation rate, whereas 
odd-parity operators are governed by  quasiparticles 
with the fastest relaxation rate. 
If $\tilde C=CT$ or $CP$, it would mean 
 Hall currents  relax more rapidly than electric currents. 
This is a situation that actually occurs in the vicinity 
of a superconducting transition, where  the  presence 
of a pairing field introduces scattering that is 
sensitive to $\tilde C = CP$.\cite{hikami}  In our case however, 
optical Hall measurements\cite{drew} already rule this 
possibility out. 
The operators $C$ and $CPT$ differ only in a spin-flip 
and for spin-less properties, they are essentially indistinguishable. 
We shall chose 
$\tilde C = C$, but our arguments are  readily modified to accommodate 
the 
alternative choice $\tilde C = CPT$. 
 
Second, 
we must be careful about the literal interpretation 
of Eq. \ref{goit}.  The charge conjugation operator does not commute 
with charge, so a  term of the type 
we are proposing  can not exist in an environment of 
unbroken symmetry. This faces us with a dilemma, 
for we  know that on a macroscopic 
scale,   the normal state 
of the cuprates has no broken symmetry. 
Our symmetry analysis forces us to conclude 
that {\em if} the Hall and electric currents have 
different decay rates, then the electrons must 
perceive their local environment as having developed a 
broken symmetry. We are thus forced to conclude that 
there must be some kind of low-energy, charge carrying 
excitation whose fluctuations create a {\em local} environment 
which 
is symmetry-broken on a time-scale $\tau \gtrsim \tau_{tr}$. 
From this point-of-view, 
(\ref{goit}) should be regarded as a mean-field assumption 
that the charge 
carrying fluctuations  produced by this environment are 
sufficiently slow, that the vertex corrections can be captured 
by an anomalous 
self-energy. We shall return to this point in the final section.

If indeed scattering is sensitive 
to the charge conjugation parity, 
then we 
might expect other transport currents to reflect these two transport 
relaxation times. Neutral currents, such as the 
thermal or thermo-electric current are also even under the charge conjugation 
operator, thus we expect that their relaxation will be governed by the 
same $T^2$ relaxation rate as the Hall current.  Circumstantial 
support for this idea is obtained from the thermopower of optimally 
doped compounds. 
 
Thermal and electric transport is normally described in terms of four 
fundamental transport tensors\cite{onsager} 
\begin{equation} 
\begin{array}{cc} 
\vec j_e &= \underline{\sigma} \vec E +  \underline{\beta} \vec 
\nabla T  \; , \\ 
\vec j_t &= \underline{\gamma}\vec E + \underline{\zeta } \vec \nabla 
T \; . 
\end{array} 
\end{equation} 
These tensors are directly linked to microscopic charge and thermal 
current fluctuations via Kubo formulae. Table I compares the leading 
temperature dependences of the various transport tensors measured in 
the optimally doped cuprates with a series of calculations we now 
describe.  The thermo-electric conductivity $\underline{\beta}$, 
determined from the conductivity and Seebeck coefficients, $\underline 
S$, $\underline{\beta} = -\underline{ \sigma} \underline{ S}$ has a 
particularly revealing temperature dependence.  In a na\"\i ve 
relaxation-time treatment, the temperature dependence of $\beta$ is 
directly related to the relevant quasiparticle relaxation rate 
$\tau_{TE}^{-1}$ according to\cite{abrikosov} 
\begin{equation} 
\beta =  -\left( \pi^2 k_B \over 3 e  \right) \left( 
k_B T \over \epsilon_F \right) 
{ n e ^2 \over m }\tau_{TE}\; , 
\end{equation} 
where $\epsilon_F$ is the Fermi energy.  Combining this with the 
electrical conductivity, $\sigma = {ne^2 \over m }\tau_{tr}$, the 
dimension-less thermopower is then 
\begin{equation} 
\tilde S = {eS \over k_B} = \left( 
{\tau_{TE} \over \tau_{tr} }  \right) 
\left( {\pi^2 \over 3}   \right) 
\left( {k_BT \over \epsilon_F}   \right) \; . 
\end{equation} 
In optimally doped compounds\cite{obertelli}, 
the thermopower contains an unusual constant part, $\tilde S\approx 
\tilde S_o - bT$ where $\tilde S_o\sim 0.1$, which indicates that 
\begin{equation} 
\tau^{-1}_{TE} = {T^2 / W_{th}} \; , 
\end{equation} 
is a factor $T/\eta W_{th}$ smaller than the transport relaxation rate, 
where 
$W_{th} = ( 3 \tilde S_o/ \pi^2 \eta ) \epsilon_F 
\sim {\epsilon_F /10}$. 
The comparable size and temperature dependence of $\tau^{-1}_{TE}$ and $ \tau^{-1}_H$ 
suggest that the same type of quasiparticle carries both the Hall current and the 
thermo-current. 
 
\section{Charge Conjugation Eigenstates and the Derivation of the Transport 
Equation} 
 
Having argued the case for a scattering mechanism which is sensitive 
to charge-conjugation parity, we now establish the effect on the 
transport properties. We need to 
express  currents in terms of 
charge-conjugation eigenstates rather than the charge eigenstates, 
electrons (and holes) that we are 
familiar with.  This  amounts to a change of basis. 
While the 
self-energy of the carriers will now be diagonal in this new basis, 
external applied fields will, in general, be able to 
inter-convert the two charge conjugation eigenstates. 
 
The eigenstates of the charge 
conjugation operator defined in Eq.~\ref{cpt} may be written 
\begin{equation} 
\begin{array}{rcl} 
\displaystyle 
a_{{\rm \vec p} \sigma} &=& \displaystyle 
{ 
%\textstyle 
 {1 \over \sqrt {2}}} 
[ \psi_{{\rm \vec p} 
\sigma}+ \sigma \psi{^{\dag}}_{{\rm \vec p}^* -\sigma} ], 
\qquad(C=+1) \\ 
\displaystyle 
b_{{\rm \vec p} \sigma} &=& \displaystyle 
{ 
%\textstyle 
 {1 \over \mbox{i}\sqrt {2}} } 
[ \psi_{{\rm \vec p} 
\sigma}- \sigma\psi{^{\dag}}_{{\rm \vec p}^* -\sigma} ]. 
\qquad(C=-1) 
\end{array} 
\label{defab} 
\end{equation} 
Fermions which are eigenstates of the charge conjugation operator 
were first introduced by Majorana over sixty years ago and, 
for this reason, are called, 
``Majorana'' fermions.\cite{majorana} When we construct a Majorana 
representation of the Fermi surface we are, in essence, folding 
the quasiparticle states inside the Fermi surface to the outside. The momenta 
of all Majorana fermions 
are restricted to the outside of the Fermi surface, for 
particles inside the Fermi surface are the anti-particles of those 
outside. 
Our central hypothesis is that {\sl quasiparticle-states of opposite 
charge conjugation parity have different relaxation rates}. 
 
\begin{figure}[tb] 
% ********   This is for two columns 
\epsfxsize=3.7in 
% ***********For one column  ******************** 
%\epsfysize=5.5in 
% ***********************************8 
\centerline{\epsfbox{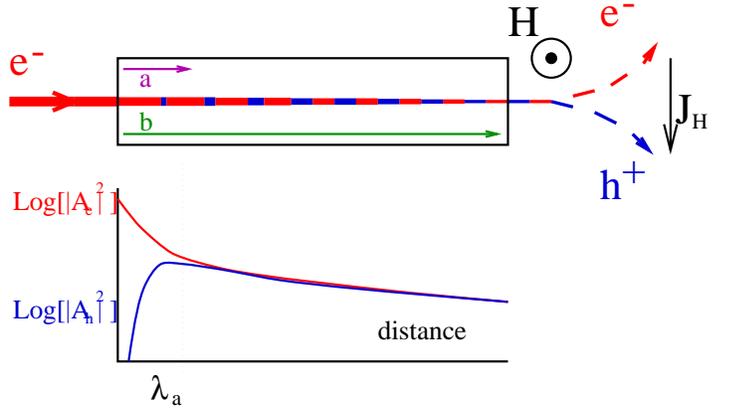}} 
\vskip 0.3truein 
\protect\caption{A simple picture of transport where relaxation processes 
are sensitive to charge conjugation parity. An electron current converts 
into an equal mixture of $a$ and $b$ charge conjugation eigenstates. One 
parity decays rapidly leaving a neutral current with equal numbers of 
electrons and holes moving together. In a magnetic field this neutral 
current can still induce a Hall voltage. 
\label{inject}} 
\end{figure}

To illustrate how this 
leads to distinct relaxation rates controlling the Hall and electric 
currents, consider the thought experiment illustrated in Fig.~\ref{inject}. 
Imagine a flux of electrons injected 
into a block of cuprate metal. Inside the metal 
we must consider these electrons to be a linear combination 
of $a$ and $b$ states 
\begin{equation} 
\psi^\dagger_{{\rm \vec p}\sigma} \longrightarrow 
a^\dagger_{{\rm \vec p}\sigma} + i b^\dagger_{{\rm \vec p} \sigma} \; . 
\end{equation} 
As the $a$ and $b$ quasiparticles propagate through the metal one species, 
say $a$, rapidly decays and becomes incoherent. At distances greater than 
the mean free path of the $a$ quasiparticles  only 
$b$ particles remain. 
The residual  current is  neutral, with equal numbers of 
electrons and holes moving at the same velocity. Charge 
transport is thus controlled by the short mean free path. 
However, should 
the $b$ particle flux move through a region of finite magnetic field, the 
Lorentz force will deflect the electron and hole components in {\em opposing} 
tangential directions. Since this neutral current 
can generate a 
finite Hall current,  the Hall response is controlled by the longer 
mean free path. From this simple thought experiment we may make the following 
general observations which will apply to this phenomenology 
\begin{itemize} 
\item the physics is insensitive to which $C$ eigenstate 
decays faster, 
\item to cleanly disentangle the separate lifetimes requires that $\Gamma_f 
\gg \Gamma_s$, 
\item were we to repeat the thought experiment choosing 
instead  eigenstates of $\tilde C= CP$ or $CT$ --- i.e. 
linear combinations of 
electrons and holes moving with opposite velocities --- then
in this case the long-lived quasiparticle carries electric current. However, 
application of a magnetic now deflects the electron 
and hole components in the same direction, so there 
is no Hall response.  So in this case, 
Hall currents  decay quickly and  electric 
currents decay slowly.  This is the situation 
near a superconducting phase transition, where the 
long-lived quasiparticles at the Fermi surface are eigenstates 
of $CP$.\cite{hikami} 
\end{itemize} 
 
We now develop the transport theory which follows from 
our  phenomenological assumption. We will not 
discuss  spin transport,  restricting our attention to 
 a simplified,  spin-less electron fluid . 
One may use diagrammatic 
perturbation theory to derive the conductivities 
and this is done in Appendix~B for 
completeness. However, since high order 
transport properties (such as the magneto-conductance) require the careful 
analysis of rather a large number of diagrams, we also develop a Boltzmann 
transport equation. In the limit of well defined quasi-particles near the 
Fermi surface, the results are equivalent. The derivation of the Boltzmann 
equation is again somewhat technical and may be found in Appendix~C. 
There we derive the following matrix generalization of the Boltzmann 
equation 
\begin{equation} 
\underline{\dot{\rm f}}  +  {\textstyle{1 \over 2}} 
\left\{ 
\underline{\vec{\cal V}}_{{\rm \vec p}}, 
\vec{\nabla}_{\rm R}  \underline{\rm  f} \right\}_+ 
+ {\textstyle{e \over 2}} 
\left\{ (\vec {\rm E} + 
\underline{\vec {\cal V}}_{{\rm \vec p}}\times \vec{\rm  B} ) 
\underline{\tau}_2 , 
\vec \nabla_{\rm p} \underline{\rm f} \right\}_+ 
= 
{\rm I} [\underline{g}] \; , 
\end{equation} 
where 
\begin{equation} 
\underline{\vec{\cal V}}_{\vec{p}}= {1 \over 
2}\left(\vec{v}_{\vec{p}}+\vec{v}_{\vec{p}*} \right) \underline{1} + 
{1 \over 2}\left(\vec{v}_{\vec{p}}-\vec{v}_{\vec{p}*} \right) 
\underline{\tau}_2 \; , 
\end{equation} 
and 
\begin{eqnarray} 
\underline{f}(t,\vec R,\vec p) &=& 
\left( 
\begin{array}{cc} 
\langle a^\dagger_{\vec p} a^{}_{\vec p} \rangle & 
\langle b^\dagger_{\vec p} a^{}_{\vec p} \rangle \\ 
\langle a^\dagger_{\vec p} b^{}_{\vec p} \rangle & 
\langle b^\dagger_{\vec p} b^{}_{\vec p} \rangle 
\end{array} 
\right)_{\vec R,t} \; 
\end{eqnarray} 
is  the quasiparticle density matrix. 
This matrix measures the local density of quasiparticles at the course-grained point 
$\vec R$ and its off-diagonal elements allow for the 
quantum superposition of  $a$ and $b$ particles. 
 
We see that the  left hand side of the transport equation is 
similar to a conventional Boltzmann equation: there are driving 
terms due to gradients in the distribution function 
 and due to electro-magnetic fields. The new features are the 
anti-commutators, which come from making a gradient expansion 
with matrices rather than single functions, and the presence of the 
second Pauli matrix $\underline{\tau}_2$. This is a reflection of the fact 
that the EM field couples to charge and, since $a$ and $b$ do not have well 
defined charge, they can be inter-converted by the applied field. 
 
The right-hand side of the transport equation contains the essence of our 
phenomenology: the collision integral. It is a functional of the 
departure from the equilibrium distribution 
$\underline{g} = \underline{\rm f} - \underline{\rm f}^{(0)}$. 
The simplicity of the experiments forces us to the hypothesis that 
the return to equilibrium is governed by 
two independent relaxation times---one for each of the charge conjugation 
eigenstates. This is represented by the collision integral 
\begin{equation} 
{\rm I} [\underline{g}]=-{1\over 2} 
\left\{\underline{\Gamma},\underline{g} \right\}_+=-{1\over 2} \left\{ 
\left( \begin{array}{cc} \Gamma_f & 0 \\ 0 & \Gamma_s 
\end{array} \right),\underline{g} \right\}_+ \; . 
\end{equation} 
Putting $\Gamma_f=\Gamma_s$ one recovers the usual relaxation time 
approximation of text book treatments. 
 
Our transport equations are completely general for arbitrary 
Fermi surface 
and anisotropic scattering rates but, as we have shown, 
including these features will not account 
for the products in relaxation rates 
appearing experimentally in magneto-transport. We therefore 
make the simplifying assumption that we have a cylindrical 
Fermi surface and that $\Gamma_f$ and $\Gamma_s$ are momentum independent. 
 
Setting up the transport equation under these conditions we note that, 
near the Fermi surface, we have 
$\delta p^* = \delta p + O(\delta p^2/p_F)$ where the small correction 
does not enter into the leading order (in $T/E_F$) transport coefficients. 
We may therefore write 
\begin{equation} 
\underline{\vec {\cal V}}_{\vec{p}} = \vec{v}_F \underline{1} 
 + \left({\delta\vec{p}\over m}\right) 
\underline{\tau}_2 \; , 
\end{equation} 
where $\vec{v}_F$ and $\delta\vec{p}\; (\geq 0)$ are normal to the 
Fermi surface.  For the in-plane transport properties we discuss here, 
 $\vec E$ and $\vec \nabla T$ lie  in the basal 
plane and the magnetic field 
is always perpendicular to the cuprate  layers. 
 
In the absence of applied fields, the distribution of quasiparticles is 
given by a diagonal matrix since, by construction, the Hamiltonian is 
diagonal in the basis of our charge conjugation eigenstates: 
\begin{equation} 
\underline{\rm f}^0_{\vec p} = {1\over 2} \left[ n_F(\epsilon_p) + 
n_F(-\epsilon_{p^*}) \right]\underline{1} \; .
\end{equation} 
The derivative, ${\vec \nabla}_{\vec p}\underline{\rm f}^0_{\vec p} = n_F' 
\underline{\vec {\cal V}}_{\vec p}$, is not 
purely diagonal. 
Here $n_F'$ is energy derivative of the Fermi function 
$\partial_\epsilon n_F(\epsilon_{\vec p})$. 
 
By expressing the deviation from the equilibrium distribution 
in the Pauli matrix basis 
$\underline{g}_{\vec p}=g_0({\vec p}) 
\underline{1}+{\vec g}(\vec{p})\cdot \underline{\vec \tau}$ and substituting 
into the Boltzmann equation, we find that the component $g_1$ decouples. 
The remaining components satisfy 
\begin{equation} 
\left( \underline{a}+\underline{b} \right) 
\left( \begin{array}{c} g_0(\vec p) \\ g_2(\vec p) \\ g_3 ({\vec p}) 
\end{array} \right) = c \; , 
\end{equation} 
where, with $\Gamma_\pm = (\Gamma_f \pm \Gamma_s)/2$, we have 
\begin{equation} 
\underline{a}=\left( 
\begin{array}{ccc} 
\Gamma_+ &     0    & \Gamma_- \\ 
    0    & \Gamma_+ &    0     \\ 
\Gamma_- &     0    & \Gamma_+ 
\end{array} 
\right) 
\quad \; , \quad 
\underline{b}= 
\left( 
\begin{array}{ccc} 
h_{\vec p} & w_{\vec p} &   0    \\ 
w_{\vec p} & h_{\vec p} &   0    \\ 
     0     &     0      &h_{\vec p} 
\end{array} 
\right) \; , 
\end{equation} 
\begin{equation} 
c = -n_F' \left( 
\begin{array}{c} 
e{\vec E}\cdot {\delta \vec p}/m - v_F^2 \hat{p} \cdot {\vec \nabla T}/T \\ 
ev_F{\vec E}\cdot \hat{p} - (v_F/m) \delta{\vec p} \cdot {\vec \nabla T}/T 
\\ 
0 
\end{array} 
\right) \; , 
\end{equation} 
with 
\begin{eqnarray} 
h_{\vec p} &=& v_F \hat{p} \cdot 
{\vec \nabla}T \partial_T + ({e/m})\delta {\vec p} 
\times {\vec H} \cdot {\vec \nabla}_{\vec p} \; , \\ 
w_{\vec p} &=& (\delta\vec p/m) \cdot {\vec \nabla T} \partial_T 
+e\left({\vec E} + v_F \hat{p} \times {\vec H} \right) \cdot {\vec 
\nabla}_{\vec p} \; . 
\end{eqnarray} 
 
We solve these equations using the Zener-Jones method~\cite{ziman}, 
solving order by order in the fields. We may write the solutions 
schematically as 
\begin{equation} 
g^{(n)}=\left(-\underline{a}^{-1}\underline{b} \right) g^{(n-1)} \quad , 
\quad g^{(1)}= \underline{a}^{-1} c \; . 
\end{equation} 
The first order solution is 
\begin{eqnarray} 
g_2^{(1)} &=& -{n_F' \over \Gamma_+}  \left( 
ev_F{\vec E}\cdot \hat{p} - {v_F \over mT} \delta {\vec p} \cdot {\vec \nabla T} 
\right) \; , \\ 
\left[ \begin{array}{c} 
g_0^{(1)} \\ g_3^{(1)} \end{array} \right] &=& -{n_F' \over \Gamma_f 
\Gamma_s} 
\left[ \begin{array}{c} \Gamma_+ \\ \Gamma_- \end{array} \right] 
\left( 
{e{\vec E}\cdot \delta {\vec p}\over m} - {v_F^2 \over T} \hat{p} \cdot {\vec 
\nabla T}  \right) \; . 
\end{eqnarray} 
This is sufficient to compute the lowest order transport 
coefficients. Off-diagonal conductivities come from the second-order 
solution. The 
magnetic field dependent part of $g^{(2)}$ is then 
\begin{equation} 
g_2^{(2)} = {en_F' \over m}  \left[ 
{ev_F  \hat{p} \times {\vec H} \cdot {\vec E} \over \Gamma_f \Gamma_s} 
- {v_F \delta p^2 {\hat p} 
\times \vec{H} \cdot {\vec \nabla T} \over \Gamma_+^2 mT} 
\right] , 
\end{equation} 
\begin{equation} 
\left[ \! \! \begin{array}{c} 
g_0^{(2)} \\ g_3^{(2)} \end{array} 
\! \! \right] \! \! \! = \! {-en_F' \over m 
\Gamma_f \Gamma_s} \! \! 
\left[ 
\begin{array}{ccc} 
{{\Gamma_+^2 - \Gamma_-^2} \over \Gamma_f \Gamma_s} & -1  & 0 \\ 
{2 \Gamma_+ \Gamma_- \over \Gamma_f \Gamma_s} & 0 & {\Gamma_- \over 
\Gamma_+ } 
\end{array} \right] \! \! \! \left[ 
\begin{array}{c} 
\displaystyle {\rm e \delta {\vec p} \times {\vec H} \cdot {\vec E} \over m} \\ 
\displaystyle {\rm v_F^2 \delta\vec{p} \times \vec{H} 
\cdot \vec{\nabla} T \over T} \\ 
\displaystyle {\rm v_F \delta p{\vec p} \times \vec{H} 
\cdot \vec{\nabla} T \over mT } 
\end{array} 
\right] . 
\end{equation} 
Finally we require the third order solution to obtain the 
magneto-conductivity. Assuming no temperature gradient the leading 
term in $T/E_F$ will come from 
\begin{equation} 
g^{(3)}_2 = {-e^3 n_F' v_F \over m^2 \Gamma_+} 
{\Gamma_+^2 + \Gamma_-^2 \over (\Gamma_f \Gamma_s)^2 } 
(\hat{p} \times \vec{H}) \cdot 
(\vec{E} \times \vec{H}) \; . 
\end{equation} 
 
Having solved the Boltzmann equation we determine the conductivities 
from the current response in applied fields. The electric 
and thermal currents may be written in terms of $\underline{g}$ 
as follows 
\begin{eqnarray} 
\vec{j}_e &=& \sum_{\vec p> p_F} 2ev_F \hat{p} g_2(\vec{p}) + 
2{e\over m} \delta\vec{p} g_0(\vec{p}) \; , \\ 
\vec{j}_t &=& \sum_{\vec p> p_F} 2 v_F^2 \delta \vec{p} 
g_{0}(\vec{k}) + 2 {v_F \delta p^2\over m} 
\hat{p} g_2(\vec{k}) \; . 
\end{eqnarray} 
The conductivities can then be extracted and we summarize the results 
in the second column of Table~\ref{conds}. 
 
\begin{table} 
\protect\caption{ Leading temperature dependences of 
transport coefficients compared with proposed decomposition into 
two Majorana relaxation times (${\cal L}_0$ is the Lorentz number 
$\pi^2 k_B^2 / 3 e^2$).\label{conds}} 
\begin{center} 
\begin{tabular}{ccccc} 
Cond-&Majorana&\multicolumn{3}{c}{Leading T behavior}\\ 
uctivity &Fluid 
&$\Gamma_f  \gg \Gamma_s$& & \\ 
& $\times \left({m \over n e^2} \right)$ & 
%{\displaystyle n e^2 \over \displaystyle m}$ 
%\phantom{$\biggl($}& 
$(T \gg T^2)$ & Expt. & Ref.\\ 
\tableline 
% SIGMA_XX ************************************ 
$\sigma_{\rm xx}$ 
& ${\displaystyle 2 \over \displaystyle \Gamma_f + \Gamma_s}$ 
\phantom{$\biggl($}& 
$T^{-1}$ &$ T^{-1}$ & \\ 
% SIGMA_XY ************************************ 
$\sigma_{\rm xy}$ & 
${\displaystyle \omega_c \over \displaystyle \Gamma_f \Gamma_s}$ 
\phantom{$\biggl($} 
& $T^{-3}$ & $T^{-3}$ \\ 
% DELTA SIGMA_XX ****************************** 
$\Delta\sigma_{\rm xx}$ & $-{\displaystyle\sigma_{xx}\over 
\displaystyle 2}\bigl( 
{\displaystyle {\omega_c^2\over\Gamma_s^2}+{\omega_c^2\over\Gamma_f^2} 
} 
\bigr)$ 
\phantom{$\biggl($} & $T^{-5}$ & $T^{-5}$ \\ 
% BETA_XX ************************************* 
$\beta_{\rm xx}$ & $- 
{\displaystyle eT {\cal L}_0\over \displaystyle 2 \epsilon_F} 
\left({\displaystyle 1 \over \displaystyle \Gamma_+}+ 
{\displaystyle \Gamma_+ \over \displaystyle \Gamma_s\Gamma_f} 
\right) 
$ 
% \phantom{$\biggl($} 
& $T^{-1}$ & $T^{-1}$ & \onlinecite{bxx}\\ 
% BETA_XY ************************************* 
$ \beta_{\rm xy} $ & $\beta_{\rm xx} {\displaystyle \omega_c \over 
\displaystyle \Gamma_+}$ \phantom{$\biggl($} 
& $T^{-2}$ & $T^{-3}(?)$ & \onlinecite{bxx} \\ 
% ZETA_XX ************************************* 
$\zeta_{\rm xx}$ &$ -{{\cal L}_0 \over \displaystyle 2} 
\left( {\displaystyle T \over \displaystyle \Gamma_f}+ 
{\displaystyle T \over \displaystyle \Gamma_s} \right)$ 
%\phantom{$\biggl($} 
& 
$T^{-1}$ & (?) & \onlinecite{kxx} \\ 
% ZETA_XY ************************************* 
$\zeta_{\rm xy}$ & $ \zeta_{\rm xx} {\displaystyle \omega_c 
 \over \displaystyle \Gamma_+}$ 
\phantom{$\biggl($}& 
$T^{-2}$ & $T^{-1}(?)$ 
& \onlinecite{kxx,ong2} \\ 
\end{tabular} 
\end{center} 
\end{table} 
 
First note that under the conditions $\Gamma_- = 0$ ({\it i.e.} 
$\Gamma_f = \Gamma_s = \Gamma_+$ our results recover the usual 
relaxation time approximation for isotropic metals. Away from this 
point however, we see that in the Hall conductivity $\sigma_{xy}$, 
for example, a product of different scattering times appears. In 
particular, when we identify the physically measured relaxation rates 
we see 
\begin{eqnarray} 
\Gamma_{tr} &=& \left(\Gamma_f + \Gamma_s \right)/2 \; , \\ 
\Gamma_H &=& 2 \left( \Gamma_f^{-1} + \Gamma_s^{-1} \right)^{-1} \; , 
\end{eqnarray} 
(i.e. relaxation rates add in $\Gamma_{tr}$ whereas lifetimes add in 
$\Gamma_H$). 
To 
address the cuprate experiments we must introduce temperature dependences of 
these scattering times. With $\Gamma_f \sim T^{-1}$ and $\Gamma_s \sim 
T^{-2}$ ({\it i.e.} with no impurity scattering terms which is 
appropriate for optimally doped single crystal YBCO) we may extract 
the leading low temperature behavior of all the transport 
conductivities as shown in the third column of Table~\ref{conds}. 
These we may compare directly with experiments as shown in the last 
two columns of the table. 
 
\begin{figure}[tb] 
% ***********For two column  ******************** 
\epsfxsize=3.5in 
% ***********************************8 
\epsfbox{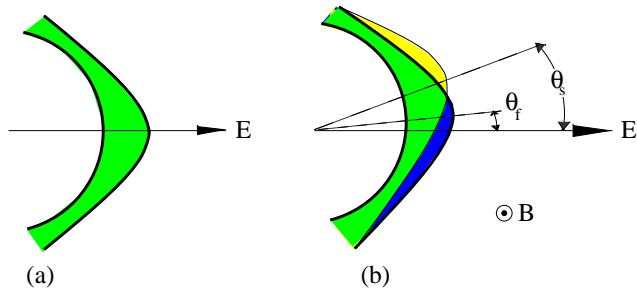} 
\protect\caption{(a) Application of field creates a mixture of 
slowly and rapidly relaxing quasiparticles. (b) Slow and fast component 
of the Majorana fluid precesses in a field, developing their 
own Hall angle  $\theta_{s,f} = \omega_c/ \Gamma_{s,f}$. 
\label{hallangles}} 
\end{figure} 
A simple physical picture of the effect of an electric field is 
provided in Fig.~\ref{hallangles}.  
When an electric field is applied, it produces an 
admixture of $C=+1$ and $C=-1$ quasiparticles whose joint relaxation 
rate $\Gamma_{tr}=\frac{1}{2}[\Gamma_s + \Gamma_f]\approx 
\frac{1}{2}\Gamma_f$ is dominated by the rapidly  relaxing 
quasiparticles.  Magnetic fields couple diagonally to the Majorana 
quasi-particles, causing fast and slow components to develop their own 
Hall angle $\theta_{s,f} = \omega_c/ \Gamma_{s,f}$ (Fig.~\ref{hallangles}).  
Since 
$\theta_s \gg \theta_f$, the Hall current is entirely dominated by the 
slow-relaxation quasiparticles. This has the effect of producing a 
finite magneto-resistance, even for isotropic Fermi surfaces: 
$\frac{\Delta \rho}{\rho} = \tan^2 \theta_H$. 
 
For completeness we also include in our table the thermal transport 
conductivities. Of these, the thermo-power and thermal conductivity 
have been most studied. 
A thermal gradient couples diagonally to the quasiparticles, so 
thermal and thermo-electric conductivities are determined by the slow 
relaxation rate.  The difference in the relaxation times of the 
electrical and thermo-electric currents then gives rise to the unique 
temperature independent component in the Seebeck coefficient $S = 
-\rho \beta \propto (T\Gamma_f / \Gamma_s)$ as we have already 
indicated. We can not compare 
our prediction of the thermal conductivity to experiment since we are 
unable to extract the large phonon contribution. Magneto-thermal 
measurements are still at an early stage but these are included in 
the table where results have been published. 
 
Thus far we have neglected spin. In the absence of a measurement of 
the ``spin conductivity'' it is not  clear which 
scattering rate will dominate this quantity. 
By choosing eigenstates of $C$ we have 
developed a phenomenology in which spin currents decay with the slow 
relaxation rate---spin-charge decoupling. 
However, if eigenstates of $CPT$ were used then in 
addition to the results already derived we would have a phenomenology 
where spin currents and electric currents decay with the fast rate. 
 
Finally, the results derived here may be extended to finite frequency 
by the replacement $\Gamma_f \rightarrow \Gamma_f(\omega)-i\omega$, 
$\Gamma_s \rightarrow \Gamma_s(\omega) -i \omega$. Detailed fits to 
experiment require a knowledge of the frequency dependence of the 
relaxation rates and therefore a microscopic model. However, at low frequencies 
we may assume that relaxation rates are frequency independent and 
so optical measurements provide an important check on our phenomenology. 
For the A.C. Hall conductivity $\sigma_{xy}(\omega)$ our model predicts 
\begin{eqnarray} 
\tan \theta_H(\omega)= 
{\sigma_{xx}(\omega)\over \sigma_{xy}(\omega)} = 
{ \omega_c \over 2} \left( {1\over \Gamma_f -i\omega} + {1 \over \Gamma_s 
- i\omega} \right) \;  ,\end{eqnarray} 
There is thus a slow, and a fast component to the Hall relaxation. 
At low frequencies $\omega \ll \Gamma_f$, 
\begin{eqnarray} 
\cot \theta_H(\omega) \sim 2 
\left( {\Gamma_s(T)-i\omega \over \omega_c}\right), \quad 
\ ( \Gamma_s, \ \omega \ll \Gamma_f ) 
\; .\label{stuff} 
\end{eqnarray} 
In particular $Im[\cot \theta_H(\omega)] \sim \omega/\omega_c$ is 
predicted to be temperature independent. 
This should be contrasted with the skew-scattering 
model,\cite{kotliar} where 
$\omega_c\rightarrow\omega_c^*(T)\propto{1 \over T}$ 
is renormalized and 
there is only one relaxation rate $\Gamma_s= \Gamma_f$. In 
this case,   $Im[\cot 
\theta(\omega)] \propto \omega T $ is proportional to temperature. 
 
Recent optical measurements 
on $YBCO$ films by  Drew {\it et. al.} have furnished results 
qualitatively in accord with Eq.~\ref{stuff}.\cite{newdrew} 
When the sample becomes superconducting, 
a flux-lattice pinning mode develops in the optical Hall angle, 
with roughly twice the spectral weight found  at frequencies 
below $\sim 200 {\rm cm}^{-1}$ in  the normal 
state.  Our model can account for this feature in terms of the 
additional, high frequency Hall relaxation at frequencies 
$\omega\sim \Gamma_{tr}$. If the additional spectral weight in the high 
frequency component condenses into the flux lattice response, we 
expect a doubling of the transverse spectral weight. 
 One of the 
difficulties that these measurements present us with, 
is that 
$\Gamma_s$ is only a factor of $4$ times smaller than $\Gamma_f$. 
It is not clear at present, whether this is due to disorder 
in the thin-films used, or whether this constitutes a significant 
discrepancy with our phenomenology.   It would be very interesting 
in this respect, to have a direct measurement of the temperature 
dependence of  $\Gamma_s$, which is not, at present available.

\section{\bf Microscopic Implications.} 
 
We now return to address some of the key microscopic 
issues that were skirted in the phenomenology of the previous sections. 
We have been led by symmetry arguments to propose 
an electron relaxation rate that depends on the charge-conjugation 
operator 
\begin{eqnarray} 
\underline{\Gamma} = \Gamma_1 + \Gamma_2 {\rm \hat C}, 
\end{eqnarray} 
so that the even- and odd-parity charge-conjugation eigenstates 
decay at different rates. Each time the charge-conjugation operator 
acts, the following units of charge and momentum are 
transferred from the quasiparticle 
to its environment 
\begin{eqnarray} 
\Delta Q = \pm 2e, \qquad 
\Delta \vec P = \pm 2 \vec k_F. 
\end{eqnarray} 
In other words,  we require 
a metallic environment where the electrons or holes 
can emit low energy quanta of momentum and charge. The phenomenology we 
have developed is, in essence a mean-field theory where these 
low-energy quanta are treated as condensed excitations. 
 
A closely analogous situation {\sl does} occur in the presence 
of superconducting fluctuations.  A Boguilubov quasiparticle 
at the Fermi energy 
\begin{eqnarray} 
\alpha{^{\dag}} _{\rm \vec k_F}\vert \Phi \rangle 
= \frac{1}{\sqrt{2}}(c{^{\dag}}_{\rm \vec k_F \uparrow} + 
c_{\rm -\vec k_F \downarrow})\vert \Phi \rangle 
\end{eqnarray} 
is an even-parity eigenstate of $CP$.  In the presence 
of superconducting fluctuations, the relaxation rates of 
the eigenstates of $CP$ differ.  Calculations by 
Aronov, Hikami and  Larkin (AHL)  confirm this conclusion: 
the conductivity contains an enhancement due to 
superconducting fluctuations 
\begin{eqnarray} 
\sigma_{xx} \rightarrow \sigma_{xx}(1  +  \lambda), 
\end{eqnarray} 
and the Hall angle is depressed by the same 
factor 
\begin{eqnarray} 
\theta_H \rightarrow \theta_H(1  -  \lambda), 
\end{eqnarray} 
Since the conductivity and the Hall angle satisfy optical sum rules, 
in the presence of superconducting fluctuations 
the transport relaxation rate is reduced but the Hall relaxation 
rate is enhanced, 
\begin{eqnarray} 
\Gamma_{\rm tr } &\rightarrow&\Gamma_{\rm tr } (1- \lambda),\cr 
\Gamma_{H} &\rightarrow&\Gamma_{H } (1+\lambda), 
\end{eqnarray} 
An appropriate generalization of Landau Ginsburg theory,  
where the order-parameter 
carries momentum, is the 
``Brazovskii model'', 
\cite{brazovskii} with the 
action 
\def\pbar{{  p\kern-1.1ex\raise-0.2ex\hbox{/}}} 
\def\vpbar{{  \vec p\kern-1.1ex\raise-0.2ex\hbox{/}}} 
\begin{eqnarray} 
{\cal F} &=& \int d^dx\left\{ 
\frac{a}{2}\vert \psi\vert^2 + \frac{b}{4 !}\vert \psi \vert^4 
+ {\xi_o^4\over 2}\vert\bigl[\pbar^2-(2k_F)^2 
\bigr]\Psi\vert^2 \right\},\cr 
\pbar &=& -i \hbar \vec \nabla - 2 e \vec A \; , 
\end{eqnarray} 
where the pair fluctuations 
are {\em staggered}. 
An extension of the AHL calculation to this case is expected to 
to show an {\sl enhanced} current relaxation 
rate, and a {\sl depressed} Hall relaxation rate. 
 
One of the interesting questions raised by this discussion, is 
whether there might be a microscopic link between our phenomenology  and 
spin-charge decoupling.\cite{phil} 
There are 
two fascinating links that should be mentioned here. Firstly,  
the holon in a Luttinger liquid is 
a charge carrying excitation with a definite momentum. 
Holon emission would  give rise to ``electron oscillations'': 
fluctuations between degenerate electron and hole states on 
the same side of the Fermi surface of the form 
\begin{eqnarray} 
e^- \rightleftharpoons  {\rm spinon\ state} \rightleftharpoons  h^+, 
\end{eqnarray} 
which would generate the time-scale separation that we have been 
discussing.  This would be the a direct analog of the 
processes that give rise to Kaon oscillations\cite{kaons} in 
particle physics. 
Secondly,  although the holon is charged, 
since it is not possible to assign 
a well-defined {\sl sign} to the charge of a holon, 
both the spinon and the holon 
should  be regarded as \ charge-conjugation 
eigenstates.  In other-words, spin-charge decoupling may naturally 
lead to quasiparticles that are charge-conjugation eigenstates, 
a condition we need for two relaxation time-scales. 
 
Before describing our attempts to make a passage to a more microscopic theory, 
let us summarize the key questions that have to be addressed: 
\begin{description} 
 
\item[$\bullet$]What is the  nature of the low energy charged excitations 
that mix electrons and holes, 
and how does this excitation couple to the original electron fields to produce 
the anomalous scattering described above? 
 
\item[$\bullet$]What are the dynamics of the charged excitation, 
and how is it possible to produce something with a correlation length 
that exceeds the electron mean-free path, but does not  diverge 
to macroscopic lengths, over a very wide range of temperatures? 
 
\end{description} 
 
We should like to end by sketching our attempt 
to address these questions. We have tried to link our phenomenology 
with the other non-Fermi liquid properties of the cuprate metallic state. 
Optical 
conductivity measurements show that the phase angle of the conductivity 
is almost constant up to 1eV, a feature which suggests that the electron 
self-energies have a singular energy dependence 
out to high energies.\cite{bontemps} 
The appearance of a linear transport relaxation rate around the Fermi surface 
of the cuprates has been used to argue that the electron-self energy 
has a 
``marginal''  form\cite{fivefriends} 
\begin{eqnarray} 
\Sigma(\omega) \propto \omega {\rm ln} \biggl[ 
{\rm max}(\omega, T)\biggr] \; ,
\end{eqnarray} 
where ${\rm max}(\omega , T) $ is an analytic function 
with asymptotes given by $\omega$, or $T$, which-ever is bigger. 
A self-energy with this particular structure is 
unusual even in the context of non-Fermi liquid models. 
For example, in a 1D Tomonaga-Luttinger 
liquid,  the self-energy has the 
form $\Sigma(\omega) \sim \omega^\alpha$, where $\alpha$ depends on 
the interactions. 
 
\begin{figure}[tb] 
% ********   This is for two columns 
\epsfxsize=3.7in 
% ***********For one column  ******************** 
%\epsfysize=5.5in 
% ***********************************8 
\centerline{\epsfbox{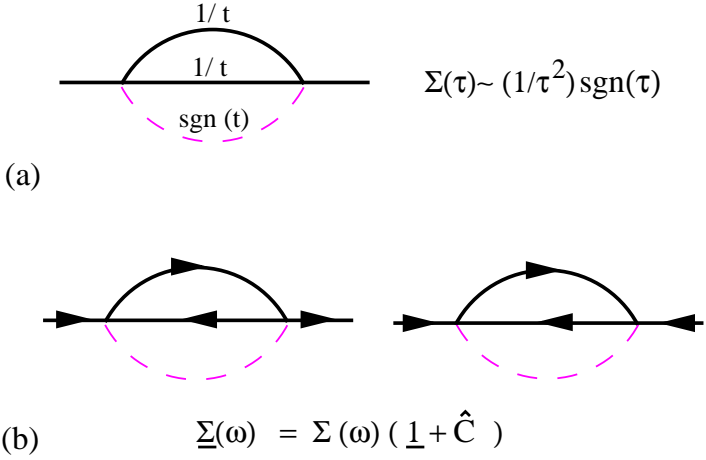}} 
\vskip 0.3truein 
\protect\caption{(a) Three-body bound-state interpretation of the 
marginal self-energy; (b) in order that the bound-state splits 
the degeneracy between the electric and Hall relaxation time, 
it must be a state of definite charge conjugation symmetry, which 
then generates 
normal and anomalous components of the self-energy of equal weight. 
} 
\label{Fig5} 
\end{figure}

There is a  way in which the two-relaxation time scenario 
can be unified with the marginal Fermi liquid picture. 
Let us suppose that the marginal 
scattering is  highly 
retarded, so that  the momentum-dependence 
of the self-energy can be ignored in our discussion. 
Since the marginal self-energy is scale invariant, we 
expect it to obey a simple power-law in the time domain. 
By  power-counting, it follows that 
\begin{eqnarray} 
\Sigma(t) \propto {\rm 1 \over t^2 } {\rm sgn}(t) \; .
\end{eqnarray} 
We identify $\Sigma(t)$ as the time-dependence of the three-body 
amplitude for the intermediate 
{\sl three-body} state formed in electron-electron scattering. 
The local propagator of electrons in the Fermi sea has the form 
$G(t) \sim {\rm 1 \over t}$.  In a Fermi liquid 
$\Sigma(t) \sim [G(t)]^3\sim t^{-3}$.   In a marginal Fermi liquid, 
the intermediate three-body state can be written in the 
suggestive form 
\begin{eqnarray} 
\Sigma(\tau) = \lambda^2 G(\tau)^2 \left[ {1\over 2} {\rm 
sgn} (\tau)\right].\label{marginalhoho} 
\end{eqnarray} 
We interpret the 
additional term on the right-hand-side 
as a zero-energy Fermionic 
excitation, without internal dynamics, as shown in Fig.~\ref{Fig5}(a). 
This hypothetical excitation may be represented 
by a single Fermi field $\Phi$, with propagator 
\begin{eqnarray} 
G^{(0)}_{\Phi}(\vec x - \vec x', \tau) = {1 \over 2} 
\delta(\vec x - \vec x') {\rm sgn}(\tau). 
\end{eqnarray} 
In the frequency domain 
\begin{eqnarray} 
G^{0}_{\Phi}(\vec k , \omega) = {1 \over \omega} \; .
\end{eqnarray} 
The possibility that such a  three-body excitation 
might drive marginal Fermi liquid behavior was first 
considered by Ruckenstein and Varma. \cite{ruckenstein} 
In its original form, the three-body 
hypothesis did 
not explain why such a resonance is pinned to the Fermi energy. 
Generically, a fermionic mode 
at the Fermi energy will develop a self-energy 
that tends to repel it  from the Fermi energy, 
eliminating the scale-invariance which 
is fundamental to the marginal scattering. 
 
The arguments of the previous sections suggest an intriguing 
unified solution to this problem.   Experiment 
tells us that the  marginal scattering rate 
selectively relaxes the electrical current: we have argued that 
 to do this, the marginal scattering 
must occur in a channel with a definite charge conjugation symmetry. 
If this is the case, then the hypothetical three-body bound-state must 
have a definite charge-conjugation parity. We are free to take 
this parity to be positive, so that 
\begin{eqnarray} 
\Phi{^{\dag}}(x)= C{^{\dag}}\Phi(x)C= + \Phi(x) \; .
\end{eqnarray} 
Such a state would immediately open 
up a decay mode of the form 
\begin{eqnarray} 
{\rm e}^- \rightleftharpoons  ({\rm e}^-  {\rm h}^+  \Phi) 
\rightleftharpoons  {\rm h}^+ , 
\end{eqnarray} 
which would then cause the $C=+1$ eigenstates of the electron 
fluid $\frac{1}{\sqrt 2}({\rm e}^-+ {\rm h}^+) $ to decay more 
rapidly than their $C=-1$ counterparts.  In the language of 
spin-charge decoupling mentioned above, the intermediate state 
might be identified with the spinon. 
 
This hypothetical $\Phi$ fermion that mediates electron oscillations 
is  automatically particle-hole symmetric. 
Furthermore, the 
momentum-independent part of its self-energy $\Sigma_{\Phi}(\omega)$ 
is necessarily an odd-function of 
frequency, $\Sigma_{\Phi}(\omega)= -\Sigma_{\Phi}(-\omega)$. 
The $\Phi$ fermion couples to the conduction electrons, to develop 
a self-energy  $\Sigma_\Phi(\tau) \propto 1/\tau^3$, so that 
 $Im \Sigma_\Phi(\omega) \propto 
\omega^2$ at low frequencies, which means that local decay processes 
do not broaden the sharpness of this resonance. 
The only way such a mode 
can develop a width is through the development of coherent 
coupling between different sites. In the absence of coherence, 
the momentum dependence of $\Sigma_\Phi$ can be neglected so the 
the sharp pole in its 
propagator is preserved. i.e, if 
\begin{eqnarray} 
G_{\Phi}^0({\omega})\longrightarrow G_{\Phi}({\omega})= 
{1 \over \omega - \Sigma_{\Phi}({\omega})}, 
\end{eqnarray} 
then 
\begin{eqnarray} 
{1 \over \pi} 
{\rm Im}\bigl[ G_{\Phi}({\omega})\bigr] 
= Z \delta({\omega}) + ({\rm background})\; , 
\end{eqnarray} 
where $Z^{-1}= [ 1 - \delta_{{\omega}}\Sigma_{\Phi}({\omega})]\vert_0$.

A three-body bound-state with definite charge-conjugation 
symmetry arises in 
the two-channel Kondo model, where it is a consequence of channel-symmetry. 
\cite{colemanioffetsvelik} 
In our case, the singular scattering derives from 
a spontaneous formation of these bound-states, presumably at energies 
around $1.0$eV, where the phase angle of the optical conductivity 
becomes constant.  We can  use this idea to make the marginal 
self-energy more precise.  Let us generalize the 
form of Eq.~\ref{marginalhoho} to finite temperatures, writing 
\begin{eqnarray} 
G(\tau) = {\pi \rho T \over \sin (\pi T \tau)}\; , 
\end{eqnarray} 
where $\rho$ is the density of states. 
Fourier transforming this we may derive 
the frequency dependent self-energy as follows 
\begin{eqnarray} 
\Sigma( i \omega_n) = {i \over 2 \beta} \int_{\Lambda^{-1}}^{\beta -\Lambda^{-1}} 
\sin{\omega_n \tau} 
\left[{\lambda \pi \rho T \over \sin (\pi T \tau)}\right]^2 
, 
\end{eqnarray} 
where $\Lambda$ is a cut-off. 
Carrying out the integral and 
analytically extending to real frequencies, we obtain 
\begin{eqnarray} 
{1 \over \tilde \lambda^2}\Sigma(\omega- i \delta) 
= 
- \omega \left[ 
{\rm ln}{ \tilde \Lambda \over T} 
- \Psi\left( 
1 + {i\omega  \over 2\pi  T} 
\right)\right]+i \pi T\; , 
\end{eqnarray} 
where $\tilde \Lambda = \Lambda e^{1-C}/2 \pi $ and 
$\tilde \lambda = \lambda \rho$. 
This function has imaginary part 
\begin{eqnarray} 
\Lambda (\omega) = {\lambda^2 \omega \over 2} 
\coth \bigl[ 
{\omega \over 2 T} 
\bigr]\; . 
\end{eqnarray} 
Scale invariant marginal Fermi liquid behavior thus has a unique 
cross-over from high to low frequencies. 
 
Let us now discuss the hypothetical interaction vertex 
illustrated in Fig.~\ref{Fig5}.  How can a three-body bound-state 
interact selectively with only one component of the Fermi sea? 
To bring out the gauge invariance of the problem, 
we need a slightly more general definition of the Majorana fermions 
around the Fermi surface. If we write the conduction electrons 
in terms of the following  matrix spinor 
\begin{eqnarray} 
\Psi_{\rm \vec p} = \left[ 
\begin{array}{rcl} 
\psi_{\rm \vec p\uparrow} & -\psi_{\rm -\vec p^*\downarrow}\cr 
\psi{^{\dag}}_{\rm -\vec p\downarrow} & \psi{^{\dag}}_{\rm \vec p^*\uparrow} 
\end{array} 
\right]\; , 
\end{eqnarray} 
then we can make a  decomposition into Majorana fermions as follows 
\begin{eqnarray} 
\Psi_{\rm \vec p} = {1 \over \sqrt{2}} 
(\psi^0_{\rm \vec p} 
+ i \vec \psi_{\rm \vec p}\cdot \vec \tau ). 
\end{eqnarray} 
We can 
define coarsely-localized Majorana fermions by taking the 
Fourier transform of these quantities 
\begin{eqnarray} 
\psi^{(a)}(x) = \sum _{\{\rm\vec p\} }\psi^{(a)}_{\rm\vec p} 
e^{i \rm\vec p\cdot \vec x}\; , 
\end{eqnarray} 
where the momentum sum is taken within a narrow shell 
around the exterior of the  Fermi surface.

The existence of a Majorana bound-state, amounts to 
the assumption that three-body correlations 
are governed by the following 
contractions 
\def\joinreld{\mathrel{\mkern-10mu}} 
\def\joinrelde{\mathrel{\mkern-11mu}} 
\def\joinrel{\mathrel{\mkern-9mu}} 
\def\joinrelw{\mathrel{\mkern-6mu}} 
\def\joinrelx{\mathrel{\mkern-4mu}} 
\def\joinrelxx{\mathrel{\mkern-2mu}} 
\def\joinrelxz{\mathrel{\mkern-12mu}} 
\def\relbd{\mathrel{{\bf\smash{{\phantom- \above1pt \phantom- 
}}}}} 
\def\ltdash{\raise-1.8pt\hbox{$\scriptscriptstyle |$}} 
\begin{eqnarray} 
\mathrel{\mathop{{ 
\psi^{(1)}(x) 
\psi^{(2)}(x) 
\psi^{(3)}(x) 
}}^{ 
\displaystyle 
\ltdash 
\mathrel{\mkern-10mu} 
\relbd\joinrelde 
\relbd\joinrelde 
\relbd\joinrelde\relbd\joinrelde 
\relbd\joinrelxz 
\relbd\joinrelxz 
\relbd\joinrelxz 
\relbd\joinrelxz 
\ltdash 
\joinrel 
\relbd\joinrelde 
\relbd\joinrelde 
\relbd\joinrelde\relbd\joinrelde 
\relbd\joinreld 
\relbd\joinreld 
\relbd\mathrel{\mkern-9.5mu}\ltdash}} 
= \lambda \Phi(x)\; . 
\end{eqnarray} 
If we want to be more general, we can write 
 
\begin{eqnarray} 
\mathrel{\mathop{{ 
\psi^{(a)}(x) 
\psi^{(b)}(x) 
\psi^{(c)}(x) 
}}^{ 
\displaystyle 
\ltdash 
\mathrel{\mkern-10mu} 
\relbd\joinrelde 
\relbd\joinrelde 
\relbd\joinrelde\relbd\joinrelde 
\relbd\joinrelxz 
\relbd\joinrelxz 
\relbd\joinrelxz 
\relbd\joinrelxz 
\ltdash 
\joinrel 
\relbd\joinrelde 
\relbd\joinrelde 
\relbd\joinrelde\relbd\joinrelde 
\relbd\joinreld 
\relbd\joinreld 
\relbd\mathrel{\mkern-9.5mu}\ltdash}} 
= \lambda \epsilon^{abcd} g_d(x) \Phi(x)\; , 
\end{eqnarray} 
where  $g_d(x)$ is a four-component  vector that 
describes the wave-function of the bound-state. 
Similar types of three-body correlation have been 
considered in the context of odd-frequency 
pairing correlations.\cite{threebody,balatsky} 
 
These considerations motivate us to consider 
interaction vertices of the form 
\begin{eqnarray} 
H_I = \sum_{x} \epsilon^{abcd} g_d(x) \Phi(x) \psi^{(a)}(x) 
\psi^{(b)}(x) 
\psi^{(c)}(x) \; ,
\end{eqnarray} 
as a possible common origin of marginal Fermi liquid 
behavior and two relaxation times. The quantity $g_d(x)$ 
is to be regarded as a slowly varying order parameter 
that describes the collective emission of charged 
excitations.  If we rotate to a basis where 
$g=\frac{1}{6}(\lambda,0,0,0)$, locally 
the interaction takes the form 
\begin{eqnarray} 
H_I = \lambda\sum_{x} \Phi(x) \psi^{(1)}(x) 
\psi^{(2)}(x) 
\psi^{(3)}(x)\; , 
\end{eqnarray} 
so that the three ``vector'' components of the conduction 
sea develop a marginal self-energy, whereas the one remaining 
component $\psi^{(0)}(x)$ would not couple to the bound-state, 
thereby preserving a more conventional Fermi liquid self-energy. 
In order for this picture to work, it is necessary to assume 
that the three-body wave-function, $g(x)$ has a coherence length 
that is large compared with electron mean-free paths. 
One of 
the interesting challenges, is to see whether the optical 
conductivity and Hall conductivity predicted by such a 
phenomenology is in accord with experiment. 
 
At present, these ideas are incomplete and under 
closer scrutiny they suffer from a number 
of difficulties.
In particular, 
\begin{enumerate} 
\item The description is incomplete without a Lagrangian 
for the three-body amplitude, $g(x)$.  It is difficult to 
see how to construct such a Lagrangian in a fashion that 
will preserve both charge and momentum conservation. 
 
\item What is the nature of the charged excitation represented 
by $g(x)$?  It is tempting to  identify this object 
with the ``holons'' entering into Anderson's Luttinger liquid 
scenario. 
 
\item Why does the boson represented by $g(x)$ not short-circuit 
the conductivity at low temperatures, like a super-current? 
 
\item  If the three-body bound-state is 
a local object, then  what possible aspect of the local 
quantum chemistry could give rise to it?  Contrariwise, 
if the bound-state 
is rather extended in real space, then what aspect of the Fermi 
surface would give rise to its spontaneous formation? 
 
\end{enumerate}

Though we  are clearly far from a microscopic explanation of the 
two-relaxation times, we expect certain broad features of our 
discussion to remain robust. 
We have demonstrated that experimental results, combined 
with the use of the transverse optical sum rule, 
strongly support the presence of two independent 
relaxation time scales for the Hall and electric current. 
This kind of separation between relaxation rates, 
if confirmed, indicates that the underlying quasiparticles 
must be states of definite charge conjugation parity. 
Such a separation can not occur 
without the presence of some, as yet unidentified, low energy 
excitation that carries both current {\sl and momentum}. 
 
We have discussed  the possible origin of this separation. 
One possibility is the presence of staggered superconducting 
fluctuations.  Another, not necessarily unrelated idea, 
is the possibility that what we are seeing here is some form 
of spin-charge separation. 
In this case, the low-lying 
charge excitations of definite charge conjugation 
parity might be identified with the holons of a Luttinger liquid. 
Finally, we have described 
how the presence of a zero-energy fermionic mode with definite 
charge conjugation symmetry may provide a way to unify these 
ideas with the idea of a marginal Fermi liquid. 
These ideas may not be mutually exclusive. 
Careful experiment, especially more accurate 
optical conductivity measurements clearly have a central  role to play 
in elucidating these issues. 
 
We should like to thank E. Abrahams, P. W. Anderson,  H. D. Drew, 
D. E. Khmelnitskii, G. L. Lonzarich and 
Ph. Ong for discussions related to this work. 
This research was supported by 
travel funds from NATO grant CRG 940040 (AMT \& PC), NSF grants 
DMR-93-12138 (PC) and DM-92-21907 and a travel fellowship from the 
Royal Society (AJS). 
 
\appendix 
 
\section{Diagrammatics for the Hall effect} 
 
This section summarizes the 
conventional calculation of the Hall conductivity, 
following the approach  of Voruganti et al.\cite{voruganti} 
Throughout, we  assume an 
$s$-wave 
scattering potential which avoids 
the need to consider vertex corrections. 
 
For spin-less electrons with  dispersion $\epsilon(\vec p)\equiv 
\epsilon_{\vec p}$,  the Hamiltonian in the presence of a vector 
potential is given by 
\begin{equation} 
H = \sum_{\rm \vec p} 
 \psi^\dagger_{\rm \vec p}\epsilon[\rm \vec p - e \vec A(x)] 
\psi_{\rm \vec p} \; . 
\end{equation} 
where $\vec x = i \nabla_{\vec p}$ 
is the position operator and 
$\vec A$ is an external vector potential that 
varies slowly on macroscopic length-scales. 
The field dependent part of the action, $S_A$, can be 
systematically in powers of $\vec A$  as follows 
\begin{eqnarray} 
{\cal S}_A&=& \sum_{n=1}^{\infty} 
{(-e)^n \over n!} \epsilon^{\alpha_1 \cdots \alpha_n}_{\vec p} 
\biggl(A^{\alpha_1}_{{\vec q}_1} 
\cdots A^{\alpha_n}_{{\vec q}_n}\biggr) 
\psi^\dagger_{{\vec p^+}} 
\psi^{}_{{\vec p^-}} 
\nonumber \cr 
&=& e\epsilon^\alpha_{\vec p} A^\alpha_{\vec q} 
\psi^\dagger_{{\vec p}+{{\vec q} \over 2}} 
\psi^{}_{{\vec p}-{{\vec q}\over 2}} 
\nonumber \cr 
&& 
+ {e^2 \over 2} \epsilon^{\alpha\beta}_{\vec p} 
A^\alpha_{\vec k}  A^\beta_{\vec q} 
\psi^\dagger_{{\vec p}+{{\vec k} \over 2} + {{\vec q} \over 2}} 
\psi^{}_{{\vec p}-{{\vec k} \over 2} - {{\vec q} \over 2}} + \cdots 
\end{eqnarray} 
where repeated 4-momenta and indices are summed over, 
$\vec p^{\pm} = \vec p \pm \sum_j \vec q_j/2$ and 
\begin{equation} 
\epsilon^{\alpha_1 \cdots \alpha_n}_{\rm \vec p} = 
 {\nabla _{ p^\alpha_1 \cdots  p^\alpha_n} 
} \epsilon(\rm \vec p)\; . 
\end{equation} 
The current operator is determined from 
\begin{equation} 
J^\alpha_q=-{1\over V} {\partial {\cal S}_A \over 
\partial A^\alpha_{-q}}. 
\end{equation} 
We wish to determine the current response to a 
uniform electric and  static magnetic field. The vector potential 
can be divided up into two components 
\begin{eqnarray} 
\vec A(\vec r, t) = \vec a^E (\vec r, t) +\vec a^H(\vec r) 
\end{eqnarray} 
where the time-dependent term gives rise to the electric 
field $\vec E = -i \nu \vec a^E_\nu$,  and the static 
term $\vec a^H$ 
determines the magnetic field, 
$\vec H = -i \vec q \times \vec a^H_{\vec q}$. 
For simplicity, we shall concentrate on 
a single frequency $\nu$,  and momentum $\pm \vec q$, 
where $\vec q$ 
 $\vec q$ is ultimately set to zero. 
The electrical and Hall conductivity are obtained by determining the 
current response as a power series 
in $\vec A$. Gauge invariance 
seriously reduces the numbers of diagrams which contribute to the 
conductivity at a given  order(see Ref.~\onlinecite{voruganti}). 
 
\begin{figure}[tb] 
% ********   This is for two columns 
\epsfxsize=3.25in 
% ***********For one column  ******************** 
%\epsfysize=5.5in 
% ***********************************8 
\centerline{\epsfbox{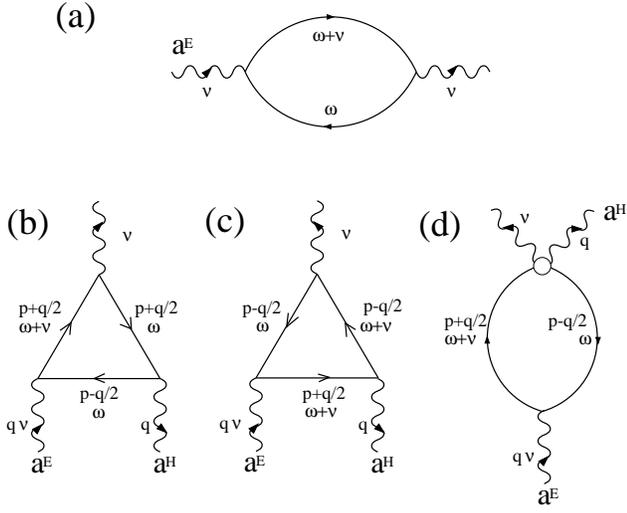}} 
\vskip 0.3truein 
\protect\caption{ 
(a) The sole diagram contributing to the electrical conductivity while 
(b)(c) and (d) show the terms contributing to the Hall conductivity. 
\label{diags}} 
\end{figure}

 To first order in $\vec A$, 
we expect currents proportional to $\vec E$ 
and $\vec H$: 
\begin{equation} 
\vec J = \sigma \vec E +\chi \vec \nabla \times \vec H \; ,
\end{equation} 
which 
correspond to diagrams proportional to $-i\nu \vec a^E_{q, \nu}$ and 
$q^2 \vec a^H_q$ . 
For the conductivity we may limit our attention to diagrams 
proportional to $i\nu a^E$. The only frequency 
dependent diagram,  is the ``bubble'' diagram,  illustrated 
in Fig.~\ref{diags}a. 
The conductivity is thus given by 
\begin{equation} 
\sigma_{xx}(\nu) = {e^2\over -i \nu} \sum_{\vec p} 
(\epsilon^x_{\vec p})^2 
{\Lambda}(\vec p, \nu)\; , 
\end{equation} 
where, 
\begin{equation} 
{ \Lambda}(\vec p, i\nu_n)= 
 {1 \over \beta} \sum_{\omega_r} 
{\cal G}^{(+)}_{\vec p} 
\bigl[{\cal G}^{(-)}_{\vec p}-{\cal G}^{(+)}_{\vec p}\bigr]. 
\end{equation} 
Here we have used the notation 
${\cal G}^{(\pm)}_{\vec p}\equiv {\cal G}(\vec p, 
i \omega_r^{\pm})$ and  $\omega_r^{\pm}= \omega_r\pm \nu_n/2$. 
For frequencies far smaller than the Fermi energy, the 
Matsubara sums give rise to a function that is quite 
sharply peaked in the vicinity of the Fermi surface. 
This permits us to replace the momentum sum by an 
energy integral as follows 
\begin{eqnarray} 
\sum_{\vec p}(\epsilon^x_{\vec p})^2 
{ \Lambda}(\vec p,i\nu_n)\rightarrow \cr 
{ n\over m} \int_{-\infty}^{\infty}d \epsilon 
{1 \over \beta} \sum_{\omega_r} 
{\cal G}^{(+)}_{\epsilon} 
\bigl[{\cal G}^{(-)}_{\epsilon}-{\cal G}^{(+)}_{\epsilon}\bigr]. 
\end{eqnarray} 
where ${\cal G}_{\epsilon}(\omega_n)= [i\omega_n - \epsilon_{\vec p} + 
\frac{i}{2}\Gamma 
{\rm sgn}(\omega_n)]^{-1} $ is the electron propagator with a phenomenological 
relaxation rate $\Gamma$, and we have assumed a parabolic 
band to  replace the angular average 
as follows 
$<v_x^2> N(0)\rightarrow n/m$. 
 
Carrying out the energy integral as a contour integral then 
gives 
\begin{eqnarray} 
\sum_{\vec p}(\epsilon^x_{\vec p})^2 
{\Lambda}(\vec p, i\nu_n) 
= 
{n\over m} 
 {\nu_n \over \Gamma + \nu_n}. 
\end{eqnarray} 
where we have taken $\nu_n>0$. 
Extending this result to  real frequencies recovers the classic 
relaxation time approximation 
\begin{eqnarray} 
\sigma_{xx}(\nu+i \delta) = {ne^2 \over m} {1 \over  \Gamma - i \nu}. 
\end{eqnarray} 
 
In an isotropic system, the second-order response determines 
the Hall current, which is 
proportional to $\vec E \times \vec H$. 
In terms of the vector potential the {\sl uniform} component of 
the  Hall current must have the following form 
\footnote{We use a slightly different convention to Voruganti et al. 
By choosing an electric field with a momentum that is precisely 
opposite to the magnetic field, we can directly extract 
the uniform Hall current. This makes 
the link between the Hall current, and the momentum-dependent 
part of the velocity operator, more apparent.} 
 \begin{eqnarray} 
j_H(\nu) \propto \nu \vec a^E({-\vec q}) 
 \times \bigl[\vec q \times 
\vec a^H({\vec q})\bigr].\end{eqnarray} 
There are six second-order 
diagrams  proportional to $a^E(-q,\nu) a^H(q)$, but gauge-invariance 
tells us that all terms must cancel excepting those which depend on a product 
of both $\nu$ and $q$. 
These diagrams are illustrated in Fig.~\ref{diags}(b-d). 
We can further simplify things by directly extracting 
the leading $q$ dependence. Adding the two triangle diagrams Fig.~\ref{diags}(b-c). 
we find 
that the terms proportional to $q$ obtained by expanding the 
Greens functions cancel, and the only residual $q$ dependence comes from 
expanding the vertex 
$\epsilon^\alpha_{k\pm q/2} \rightarrow \epsilon^\alpha_k \pm q^\beta 
\epsilon^{\alpha\beta}_k/2$. The leading $q$ dependence of 
Fig.~\ref{diags}(c). 
is obtained by expanding the Green's functions in the bubble: 
${\cal G}(k\pm q/2) \rightarrow {\cal G}(k) \pm q^\beta \epsilon^\beta_k 
{\cal G}(k)^2$. The sum of these contributions is represented 
by the triangle diagrams shown in Fig.~\ref{Fig2}b. 
The evaluation 
of these two diagrams leads to 
\begin{eqnarray} 
\sigma_{xy}(\nu)= {e^3H\over \nu} \sum_{\vec k} \left( 
\epsilon^x_{\vec k} \epsilon^x_{\vec k} 
\epsilon^{yy}_{\vec k}-\epsilon^x_{\vec k} 
\epsilon^y_{\vec{k}}\epsilon^{xy}_{\vec k}\right) 
\Pi(\vec k, \nu) \label{sigmaxy} 
\end{eqnarray} 
where 
\begin{eqnarray} 
\Pi(\vec k,i \nu_r)= 
{1 \over 2\beta} \sum_{\omega_n} 
{\cal G}^{(+)}_{\vec k} 
{\cal G}^{(-)}_{\vec k} 
\left[{\cal G}^{(-)}_{\vec k} 
-{\cal G}^{(+)}_{\vec k} 
 \right]. 
\end{eqnarray} 
In the Landau gauge, where $\vec a^H = (a^H,0,0)$ and $\vec q = (0,q,0)$, 
the first term in (\ref{sigmaxy}) is obtained from the triangle 
diagrams Fig.~\ref{diags}(b-c), whereas the second 
term comes from the q-expansion of the bubble diagram Fig.~\ref{diags}(d). 
For a parabolic band, the second term vanishes, so 
we could have completely neglected the bubble diagrams. 
As in the case of the conductivity, 
the Matsubara 
summations lead to a function that is quite sharply 
peaked within a few $kT$ of 
$\epsilon(k_F)$ , so the  momentum sum only probes the 
region close to the Fermi surface and the 
momentum sum can then be replaced by an energy integral 
as follows 
\begin{eqnarray} 
\sigma_{xy}(i \nu_n) 
= 
{ne^3H\over i\nu_n m^2 }{1 \over 2\beta} \sum_r  \tilde 
\Pi( i \omega_r^+, i \omega_r^-) 
\end{eqnarray} 
and 
\begin{eqnarray} 
\tilde 
\Pi( i \omega_r^+, i \omega_r^-)= 
\int_{-\infty}^{\infty} d \epsilon 
{\cal G}^{(+)}_{\epsilon}{\cal G}^{(-)}_{\epsilon} 
\bigl[{\cal G}^{(-)}_{\epsilon}-{\cal G}^{(+)}_{\epsilon} 
 \bigr]. 
\end{eqnarray} 
where we have used the shorthand notation 
${\cal G}^{(\pm)}_{\epsilon}\equiv{\cal G}(\epsilon, i \omega_r^{\pm})$. 
Carrying out the energy integral then gives 
\begin{eqnarray} 
\tilde 
\Pi( i \omega_r^+, i \omega_r^-)= 
{\pi i \over ( \Gamma+ \nu _n)^2}[\Theta(\omega_r^+)- 
\Theta(\omega_r^-)], 
\end{eqnarray} 
so that 
\begin{eqnarray} 
\sigma_{xy}(\nu+i \delta) 
= 
{ne^2\over m }{\omega_c \over (\Gamma-i\nu)^2}, 
\end{eqnarray} 
recovering the result of Boltzmann transport theory. 
 
\section{Diagrammatics with charge-conjugation eigenstates} 
 
We now build on the experience gained in Appendix A and 
repeat the calculation using charge-conjugation 
eigenstates and a phenomenological 
scattering rate which depends on the charge conjugation parity. 
We begin by combining electron states above the Fermi 
surface with their charge-conjugation partners below it to make a 
spinor 
\begin{eqnarray} 
\Psi _{{\rm \vec p} \sigma} = 
\left( 
\begin{array}{rl} 
\psi_{{\rm \vec p} -\sigma}\cr 
\sigma \psi{^{\dag}}_{{\rm \vec p^*} \sigma} 
\end{array} 
\right) \; . 
\end{eqnarray} 
The Hamiltonian in a field is then 
\begin{equation} 
H_o = \displaystyle\sum_{{\rm |\vec p| > |\vec p_F| }, \sigma} 
\Psi{^{\dag}}_{{\rm \vec p} \sigma} 
{\rm \epsilon}_{{\rm \vec p -e \vec A \tau_3}} 
\Psi^{}_{{\rm \vec p} \sigma} \; . 
\end{equation} 
where $\tau_3$ is the third Pauli spinor. 
The momentum sum is now restricted to outside the Fermi surface 
because the hole states are already included in the spinor. 
By differentiating with respect to the vector potential, we 
may derive the corresponding current operator 
\begin{eqnarray} 
{\cal J}= e\sum_{{\rm |\vec p| > |\vec p_F| }, \sigma} 
\Psi{^{\dag}}_{{\rm \vec p} \sigma} 
\bigl[{\rm \vec v_{\rm \hat p_F}\tau_3 +\underline{m}^{-1}(\delta 
\vec p - e \vec A\tau_3) } 
\bigr] 
\Psi^{}_{{\rm \vec p} \sigma} \; . 
\end{eqnarray} 
In this basis 
the electro-magnetic field is diagonal, but the states are 
not eigenstates of the $C$.  To make the charge conjugation eigenstate, 
we make a rotation within the degenerate electron and hole states, 
writing 
\begin{eqnarray} 
{\tilde \Psi}_{\rp \sigma} = U \Psi_{\rp \sigma} 
\end{eqnarray} 
where, 
\begin{equation} 
U= \left[ 
\begin{array}{rl} 
{1 \over \sqrt{2}}&{1 \over \sqrt{2}}\cr 
{-i \over \sqrt{2}}&{i \over \sqrt{2}} 
\end{array} 
\right] \; , \qquad 
{\tilde \Psi} _{{\rm \vec p} \sigma} = 
\left( 
\begin{array}{rl} 
a_{{\rm \vec p} \sigma}\cr 
b_{{\rm \vec p} \sigma} 
\end{array} 
\right) \; , \label{transfm} 
\end{equation} 
where $a$ and $b$ are defined in Eq.~\ref{defab}. 
The transformation to the new basis is easily effected by 
noting that 
\begin{eqnarray} 
U\tau_3U{^{\dag}} = \tau_2, 
\end{eqnarray} 
where $\tau_2$ is the second Pauli matrix. 
In the new basis, the 
Hamiltonian  is thus 
\begin{equation} 
H_o = \displaystyle\sum_{{\rm |\vec p| > |\vec p_F| }, \sigma} 
{\tilde \Psi}{^{\dag}}_{{\rm \vec p} \sigma} 
{\rm \epsilon}_{{\rm \vec p -e \vec A \tau_2}} 
{\tilde \Psi}_{{\rm \vec p} \sigma} \; . 
\end{equation} 
and the current operator becomes 
\begin{eqnarray} 
{\cal J}= e\sum_{{\rm |\vec p| > |\vec p_F| }, \sigma} 
\tilde \Psi{^{\dag}}_{{\rm \vec p} \sigma} 
\bigl[{\rm \vec v_{\rm \hat p_F}\tau_2 +\underline{m}^{-1}(\delta 
\vec p - e \vec A\tau_2) } 
\bigr] 
\tilde \Psi^{}_{{\rm \vec p} \sigma} \; . 
\label{raisin} 
\end{eqnarray} 
In the absence of an 
external field, this Hamiltonian is diagonal and charge-conjugation is 
a conserved symmetry.\cite{footy} Our phenomenological assumption is 
that the leading irrelevant interactions in this Hamiltonian yield 
independent lifetimes for the $a$ and $b$ particle: 
\begin{eqnarray} 
{\cal G}^{-1}_a(\vec p, i\omega)&=& i\omega - \epsilon_{\vec p} + 
\frac{i}{2} \Gamma_f 
{\rm sign}(\omega) \; , \\ 
{\cal G}^{-1}_b(\vec p, i\omega)&=& i\omega - \epsilon_{\vec p} + 
\frac{i}{2} \Gamma_s 
{\rm sign}(\omega)  \; . 
\end{eqnarray} 
where we have arbitrarily assigned the fast relaxation rate to the
$a$ particle. 
The matrix propagator for $\tilde \Psi$ is then 
\begin{eqnarray} 
{\cal G}(\vec p, i \omega) =\left[ 
\begin{array}{rl} 
{{\cal G}_a (\vec p, i \omega)   }&{}\cr 
{}&{{\cal G}_b(\vec p, i \omega)} 
\end{array} 
\right] 
\end{eqnarray} 
In general $\Gamma_f$ and $\Gamma_s$ will be momentum and 
frequency dependent, however out of ignorance we assume that they are $p$ 
independent (so we don't have vertex corrections). 
We will also consider them to be frequency independent, 
though this is an assumption that is readily relaxed. 
 
\begin{figure}[tb] 
% ********   This is for two columns 
\epsfxsize=3.25in 
% ***********For one column  ******************** 
%\epsfysize=5.5in 
% *********************************** 
\centerline{\epsfbox{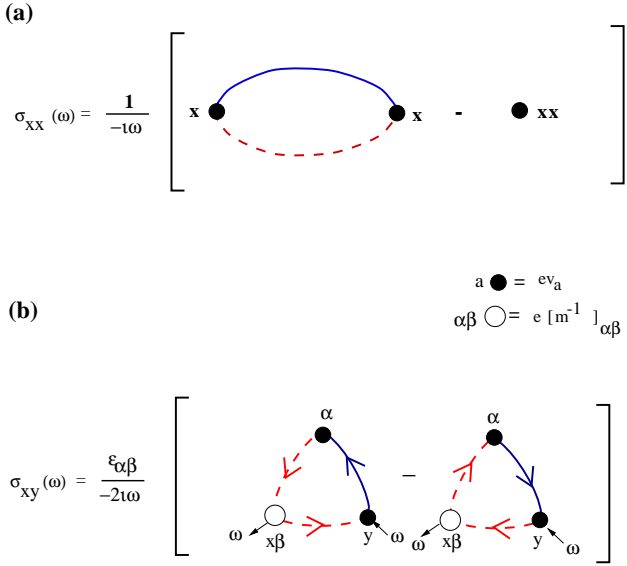}} 
\vskip 0.3truein 
\protect\caption{ 
The diagrams contributing to (a) the conductivity and (b) the 
Hall conductivity in terms of Majorana fermions. 
(There are also a set of diagrams with the two Majorana parities 
interchanged.) 
In the electrical conductivity 
both species contribute equally {\em in each diagram} and the fast relaxation 
rate will dominate. By contrast, in the Hall conductivity there is an 
{\em asymmetry} between the number of each species of Majorana 
contributing to a given diagram. While the symmetry is 
restored by adding the diagrams with the parities interchanged, adding 
the diagrams is equivalent to adding conductivities and 
allows the long lived species to ``short circuit'' the 
rapidly  decaying one in the Hall current. 
\label{abdiags}} 
\end{figure}

It is straightforward to repeat the diagrammatic approach of Appendix B in this 
matrix formalism.  Suppose for the moment we take a 
parabolic band and naively assume that bubble diagrams 
of the form Fig.~\ref{diags}(d)  may be neglected, then we can 
restrict our attention to the triangle diagrams of the type shown in 
Fig.~\ref{diags}(b) \& (c). 
From (\ref{raisin}), we see that we can quickly generalize 
the momentum-independent and momentum-dependent parts of the 
current vertices to the matrix notation by making the following 
substitutions 
\begin{eqnarray} 
\vec v_{F} &\longrightarrow &\vec v_F\otimes \tau_2,\cr 
\underline {m}^{-1}_{\rm \vec p}&\longrightarrow & 
\underline {m}^{-1}_{\rm \vec p}\otimes \underline{1}. 
\end{eqnarray} 
With these modifications, 
we find that results of Appendix A. may be generalized 
by writing 
\begin{eqnarray} 
\Lambda(\vec p, i \nu_n)&=& 
 {1 \over 2\beta} \sum_{\omega_r} 
 {\rm Tr}\left[\tau_2 
\bigl({\cal G}^{(+)}_{\vec p}-{\cal G}^{(-)}_{\vec p}\bigr) 
\tau_2 
{\cal G}^{(-)}_{\vec p}\right],\cr 
\Pi(\vec p, \nu)&=& 
{1 \over 4\beta} \sum_{\omega_n} 
{\rm Tr} 
\left\{ {\cal G}^{(+)}_{\vec p}{\cal G}^{(-)}_{\vec p} 
\tau_2\bigl[{\cal G}^{(-)}_{\vec p} 
-{\cal G}^{(+)}_{\vec p} \bigr]\tau_2 
\right\} 
\end{eqnarray} 
where we have used the shorthand notation $ 
{\cal G}^{(\pm)}_{\vec p}\equiv 
{\cal G}(\vec p, i \omega_r \pm i \nu_n/2)$. 
These expressions revert to those given in Appendix A. 
when $\Gamma_a= \Gamma_b$. A simple diagrammatic interpretation 
of both expressions can be made by noting that 
$\tau_2$ is off-diagonal in the Majorana basis. 
The  electric current operator is off-diagonal in this basis, 
and the Fermion bubble entering into the conductivity 
contains  one ``a'' and one ``b'' propagator, as shown in 
Fig.~\ref{abdiags}(a).  By contrast, the Hall conductivity involves the 
effective mass operator, which is diagonal in the ``a-b'' basis. 
The triangle diagrams entering into the Hall conductivity 
are shown in Fig.~\ref{abdiags}(b). 
 
Replacing the momentum sums by 
an energy integral, we can repeat the steps of Appendix A., 
to find 
\begin{eqnarray} 
\sum_{\vec p}(v^x_{\vec p_F})^2 
{\Lambda}(\vec p, i\nu_n) 
= {n\nu_n\over m} 
 {1 \over  \Gamma_+ +  \nu_n},\cr 
\sum_{\vec p}(v^x_{\vec p_F})^2 
\Pi(\vec p, i\nu_n)= 
{ n \over m} 
{i\nu_n \over (\Gamma_f + \nu_n)(\Gamma_s + \nu_n)} 
\end{eqnarray} 
where $\Gamma_+=\frac{1}{2}(\Gamma_f+\Gamma_s)$ and we have taken $\nu_n >0$. 
The results are then 
\begin{eqnarray} 
\sigma_{xx}(\nu) &=&{ne^2 \over m} 
{1 \over \Gamma_+ -i \nu} 
\; ,\cr 
\sigma_{xy}(\nu)& 
=& {ne^2  \over m}  {\omega_c\over 
(\Gamma_f -i \nu)(\Gamma_s-i \nu)} , 
\end{eqnarray} 
These are the same results as obtained from the phenomenological 
Boltzmann approach. 
 
The shrewd diagrammatician, will of course recognize that we have been 
too cavalier. By introducing 
terms into the scattering self-energies that  do not commute with the 
charge operator, (which is $\tau_2$ in this basis), 
we have broken gauge-invariance. In the bubble diagrams, 
the order of propagators and the charge operator is inverted, so that 
these diagrams no-longer give the same contributions as their 
triangular counterparts.  A more careful calculation which includes 
the bubble diagrams, leads to the following result 
\begin{eqnarray} 
\sigma_{xy} = \frac{1}{2}(\sigma_{\Delta}+\sigma_{O})\vec \nabla\times \vec A 
+\frac{1}{2}(\sigma_{\Delta}-\sigma_{O})\vec \nabla\cdot \vec A 
\label{gaugeinv} 
\end{eqnarray} 
where 
\begin{eqnarray} 
\sigma_{\Delta}(\nu)&=& {ne^3  \over m^2}  {1\over 
(\Gamma_f -i \nu)(\Gamma_s-i \nu)}\; ,\cr 
\sigma_{O}(\nu)&=& {ne^3  \over m^2}  {1\over 
(\Gamma_+ -i \nu)^2} , 
\end{eqnarray} 
are the contributions from the triangle and bubble diagrams, respectively. 
The  second term in ~\ref{gaugeinv} reflects the limitations of this 
kind of phenomenology. 
In a more consistent calculation, the second 
term in ~\ref{gaugeinv} would presumably enter in combination with 
the gradient of phase of the anomalous scattering vertex. 
Were we to regard the first term in \ref{gaugeinv} as the physically 
relevant component, then  the $b$ quasiparticle would still selectively short-circuit 
$\sigma_{xy}$, but the Hall angle would be reduced by a factor of two 
relative to the Boltzmann equation approach. 

\section{Derivation of the transport equation} 
 
The Boltzmann equation for charge conjugation eigenstates is 
determined from the semi-classical limit of the quantum Boltzmann 
equation. 
The derivation of the quantum Boltzmann equation (QBE) has been reviewed by 
Rammer and Smith~\cite{rammer86} and we will rely heavily 
on their results. We proceed by deriving the QBE for electrons and 
their degenerate hole states as described by Rammer and Smith but, 
since our collision term will ultimately couple these two states, we group 
them together in a matrix. The perturbations that drive the Fermi 
system out of equilibrium---EM fields and temperature gradients {\it 
etc}---couple to electrons and holes directly and 
independently. Interactions in  a 
conventional metal also preserve the electron or hole 
nature of the quasiparticles with the net result that the QBE is 
diagonal in the particle-hole basis and one solves for a single scalar 
distribution function. We wish to generalize this to the situation 
where the driving perturbations remain diagonal in the particle-hole 
basis but the interaction terms are diagonal only in the basis of charge 
conjugation eigenstates. To do this we will begin by deriving the 
driving terms in the QBE in the particle-hole basis by considering the 
collision-less system. We shall be careful though to make no assumption 
of commutivity since ultimately the collision term which we introduce 
will not be diagonal in this basis. Having obtained the driving terms 
we then rotate the equation into the basis of charge conjugation 
eigenstates where our phenomenological collision term is 
diagonal. This will then give our new transport equation from which we 
can obtain the conductivities. 
 
The QBE is usually obtained from the equation of motion of the `Keldysh' 
Green's function. We therefore consider the time evolution of the following 
matrix Green's function expressed in terms of particles and holes 
\begin{eqnarray} 
\underline{G}^K(x_1,{x_{1'}}) &\equiv& 
G^K_{ij}(x_1,x_{1'})\cr 
&=&-i\langle \left[\Psi^{\dagger}_i(x_1),\Psi_j(x_{1'}) 
\right]_{-} \rangle \; . 
\end{eqnarray} 
Here $x_1$ is shorthand for the 4-coordinate $(t_1,\vec{r}_1)$. 
We deal first with the driving term in the QBE for which we need only 
consider the collision-less regime. Here the 
difference between the equation of motion acting on the 
left and right operators in $\underline{G}^K$ may be written as 
\begin{equation} 
\left[\delta(x_1-x_{1'}) \left( i\partial_{t_1} - 
\epsilon(-i\vec{\nabla}_{\vec{r}_1}) \right) 
\underline{1} \stackrel{\otimes}{,} \underline{G}^{K} 
\right]_{-} = \underline{0} \; , 
\end{equation} 
where $\otimes$ is the convolution operator in real space. This 
equation is now coarse grained in the presence of an EM field by 
introducing the mixed representation 
\begin{equation} 
\underline{G}(X,p) = \! \int \! \! dx e^{-i\vec{r} \cdot 
\left[\underline{\vec{p}}+e\vec{A}(X) \underline{1} \right] + 
it\left[E+e\varphi(X)\right]\underline{1}} \underline{G}(X,x) , 
\end{equation} 
where $X$ is the center of mass coordinate $X=(x_1+x_{1'})/2$ 
and we Fourier transform over the relative position $x=x_1-x_{1'}$. 
To correctly transform the degenerate particle and hole states, the 
momentum $\underline{p}$ becomes a diagonal matrix 
\begin{equation} 
\underline{\vec{p}}=\left(\begin{array}{cc} \vec{p} & 0 \\ 0 & 
-\vec{p}* \end{array} \right) \; . 
\end{equation} 
This transformation has 
the combined effect of coupling the system in a gauge invariant manner 
to the electro-magnetic field $A=(\varphi,\vec{A})$ and 
expressing everything in terms of the 
kinematic momentum. Under conditions of uniform, static 
electro-magnetic fields the convolution operator can be written 
in the mixed representation using 
the gradient expansion 
\begin{eqnarray} 
G \otimes H = 
e^{ {i\over 2} 
 \left( \partial_X^G \cdot \partial_p^H - \partial_p^G \cdot \partial_X^H 
 + \partial_p^G \cdot F \cdot \partial_p^H \right) } 
G(X,p) H(X,p) 
\end{eqnarray} 
where $\partial_p^G$ denotes the derivative operator 
$(\partial_E , 
\vec{\nabla})$ that acts exclusively on $G$ and $F$ is the usual antisymmetric 
EM field tensor $\partial_\mu A^\nu - \partial_{\nu} A^\mu$. 
 
To obtain the classical limit of the QBE we simply apply the gradient 
expansion to first order to the equation of motion for 
$\underline{G}^K$. We now make the usual quasiparticle 
assumption~\cite{qp,kadanoff} that $\underline{G}^K$ is sharply peaked in 
energy $\sim \delta(\omega-\epsilon(\vec{k}))$ and we integrate the equation 
of motion over frequency. This leads then to the collision-less Boltzmann 
equation 
\begin{equation} 
{1 \over 2} \left\{ \left( 
\begin{array}{c c} 
H_{\vec{p}} & 0 \\ 
0 & H_{\vec{p}*} 
\end{array} 
\right) , \underline{G}^K (T, {\vec R}, {\vec p}) \right\}_+ = 
\underline{0} \; , 
\end{equation} 
where, with $\vec{v}_{\vec{p}} = \vec{\nabla}_{\vec{p}} \epsilon(\vec{p})$, 
we define 
\begin{equation} 
H_{\vec{p}} = \partial_T + \vec{v}_p \cdot \vec{\nabla}_{\vec{R}} 
+ e\left(\vec{E}+\vec{v}_{\vec{p}}\times \vec{B} \right) \cdot 
\vec{\nabla}_{\vec{p}} \; . 
\end{equation} 
 
We display the equations at this point to emphasis the fact that we 
have merely derived the conventional collision-less Boltzmann equation: 
the anticommutator here is trivial and we have two independent Boltzmann 
equations. 
However we now wish to express the above equation in terms 
of the charge conjugation eigenstates since our hypothesis for the 
scattering distinguishes between them. We apply the unitary 
transformation of Eq.~\ref{transfm} to rotate our transport equation 
into the basis of charge conjugation eigenstates 
\begin{equation} 
\underline{\dot{\rm f}}  +  {\textstyle{1 \over 2}} 
\left\{ 
\underline{\vec{\cal V}}_{{\rm \vec p}}, 
\vec{\nabla}_{\rm R}  \underline{\rm  f} \right\}_+ 
+ {\textstyle{e \over 2}} 
\left\{ (\vec {\rm E} + 
\underline{\vec {\cal V}}_{{\rm \vec p}}\times \vec{\rm  B} ) 
\underline{\tau}_2 , 
\vec \nabla_{\rm p} \underline{\rm f} \right\}_+ 
= 
{\rm I} [\underline{g}] \; , 
\end{equation} 
where 
\begin{eqnarray} 
\underline{\rm f}(T,\vec{R},\vec{p})&=&\int d\omega \underline{U}^{-1} 
\underline{G}^K \underline{U} = \int d\omega \langle 
\left[\tilde \Psi^{\dagger}_p \tilde \Psi^{}_p \right]_- \rangle , 
\\ \; 
\underline{\vec{\cal V}}_{\vec{p}}&=& {1 \over 
2}\left(\vec{v}_{\vec{p}}+\vec{v}_{\vec{p}*} \right) \underline{1} + 
{1 \over 2}\left(\vec{v}_{\vec{p}}-\vec{v}_{\vec{p}*} \right) 
\underline{\tau}_2 \; . 
\end{eqnarray} 
We have now added a collision integral, ${\rm I} [\underline{g}]$ 
which is a functional of the departure from equilibrium, 
$\underline{g} = \underline{\rm f} - \underline{\rm f}^{(0)}$. 
Our central hypothesis is now contained in 
this collision functional: the return to equilibrium is governed by 
two independent relaxation times---one for each of the charge conjugation 
eigenstates 
\begin{equation} 
{\rm I} [\underline{g}]={1\over 2} 
\left\{\underline{\Gamma},\underline{g} \right\}_+={1\over 2} \left\{ 
\left( \begin{array}{cc} \Gamma_f & 0 \\ 0 & \Gamma_s 
\end{array} \right),\underline{g} \right\}_+ 
\end{equation} 
Writing $\underline{\Gamma}=\Gamma_+ \underline{1} + \Gamma_- 
\underline \tau_3$ we stress again that if 
$\Gamma_-=(\Gamma_f-\Gamma_s)/2=0$  our transport 
equation is simply the usual relaxation time approximation used in text 
book treatments. The 
consequences of $\Gamma_- \neq 0$ are established in this paper.


\begin{references} 
\bibitem{linear}See for example, M. Gurvitch and A. T. Fiory, Phys. 
Rev. Lett. {\bf 59}, 1337 (1987); 
%K. Karam\'as {\it et al.}  Phys. Rev. Lett. {\bf 64},84 (1990); 
L. Forro {\it et al., ibid.} {\bf 65}, 1941 (1990). 
\bibitem{chien} T. R. Chien, Z. Z. Wang and N. P. Ong, 
Phys. Rev. Lett. {\bf 67}, 2088 (1991). 
\bibitem{phil} P. W. Anderson, Phys. Rev. Lett. {\bf 67}, 2092 (1991), 
see also Y. Ren and P. W. Anderson, Phys. Rev. B {\bf 48}, 16662 (1993). 
\bibitem{romero}D. B. Romero, Phys. Rev. B {\bf  46}, 8505 (1992).
 \bibitem{ourprl}An earlier summary of these ideas appeared in 
P. Coleman, A. J. Schofield and A. M. Tsvelik, Phys. Rev. Lett. 
{\bf 76}, 1324 (1996); cond-mat/9602001.
\bibitem{ong}J. M. Harris et al., 
Phys. Rev. Lett. {\bf 75}, 1391 (1995). 
\bibitem{onghall} N. P. Ong, Phys. Rev. B {\bf 43}, 193 (1991). 
\bibitem{ziman} J. Ziman, ``Electrons and Phonons'', Oxford University 
Press, (1960).  For a discussion of Kohler's rule, see pp. 490. For a 
discussion of the Zener-Jones method see pp. 502. 
\bibitem{tanake}I. Terasaki, Y. Sato, S. Miyamoto, S. Tajima and S. Tanaka, 
Phys. Rev. B {\bf 52}, 16246 (1995). 
\bibitem{takagi}T. Kimura , S. Miyasaka, H. Takagi et al., Phys. Rev. 
B {\bf 53}, 8733 (1996). 
\bibitem{drew}S. G. Kaplan, S. Wu, H. T. S. Lihn and D. Drew, Q. Li, 
D. B. Fenner, Julia Phillips and S. Y. Hou, Phys. Rev. Lett. {\bf 76}, 
696 (1996). 
\bibitem{drewcoleman} H. D. Drew and P. Coleman, to be published 
(1996). 
\bibitem{newdrew}H. D. Drew, S. Wu and H-T. S. Lihn, proceedings 
of this conference. 
\bibitem{carrington}A. Carrington, A. P. Mackenzie, C. T. Lin and 
J. R. Cooper, Phys. Rev. Lett. {\bf 69}, 2855 (1992). 
\bibitem{pines}B. P. Stojkovic and D. Pines, Phys. Rev. Lett {\bf 76}, 
811 (1996); cond-mat/9505005. 
\bibitem{hlubina}R. Hlubina and T. M. Rice, Phys. Rev. B {\bf 51} 
9253, (1995); cond-mat/9501086.
\bibitem{kotliar}B. G. Kotliar, A. Sengupta and C. M. Varma, Phys. Rev. 
B {\bf 53}, 3573 (1996). 
\bibitem{voruganti} P. Voruganti, A. Golubentsev and S. John, Phys. 
Rev. B {\bf 45}, 13945 (1992). 
\bibitem{hikami}Superconducting fluctuations enhance the 
conductivity, but without changing the Hall conductivity, as shown 
recently by 
A. Aronov, S. Hikami and A. I. Larkin, 
Phys. Rev. B{\bf 51}, 3880 (1995). 
This means that the conductivity is enhanced, but the 
Hall angle is suppressed by superconducting fluctuations. 
By the f-sum and t-sum rules, this means that the decay 
rates of electric 
currents  and Hall currents are reduced and enhanced 
respectively, by superconducting fluctuations. 
\bibitem{onsager}Note that $\underline{\beta}$ and 
$\underline{\gamma}$ are linked by the Onsager 
relation  $\underline{\gamma}(B) = - T \underline{\beta}^T(-B) $, where 
$B$ is the magnetic field. 
\bibitem{abrikosov} see, for example, A. A. Abrikosov, {\it 
Fundamentals of the Theory of Metals}, North Holland, p. 103 (1988). 
\bibitem{obertelli} S. D. Obertelli, J. R. Cooper and J. L. Tallon, 
Phys. Rev. B, {\bf 46}, 14928 (1992). 
\bibitem{majorana}E. Majorana, Il Nuovo Cimento, {\bf 14}, 171 (1937). 
\bibitem{bxx} $\underline{\beta}=\underline{\sigma} \underline{S} 
\approx \underline{\sigma} S_{xx}$, since the Nernst Ettinghausen 
coefficient is negligible ($|S_{xy}|/S_{xx} \ll 
\sigma_{xy}/\sigma_{xx}$): see J. A. Clayhold, A. W. Linnen Jr., F. Chien and 
C. W. Chu, Phys. Rev. B {\bf 50} 4252, (1994). 
\bibitem{kxx} $\underline{\zeta}=-\underline{\kappa}+T 
\underline{\beta}^{\rm T}(-B) \underline{\sigma} \underline{\beta}(B)$ with 
the leading temperature dependence coming from the thermal conductivity 
$\underline{\kappa}$. 
The $\kappa_{xx}$ is phonon dominated: see C. Uher, in 
{\it Physical properties of high temperature superconductors: 
vol. III} p. 159, 
editor D. M. Ginsberg, World Scientific, Singapore (1992). 
\bibitem{ong2} 
K. Krishana, J. M. Harris and N. P. Ong, 
Phys. Rev. Lett {\bf 75}, 3529 (1995). 
\bibitem{brazovskii}S. Brazovskii, Zh. Eksp. Teor Tiz. {\bf 68}, 175 (1975); 
Sov. Phys. JETP {\bf 41}, 85 (1975). 
\bibitem{kaons}M. Gell-Mann and A. Pais, Phys. Rev. {\bf 97}, 1387 (1955). 
\bibitem{bontemps}A. El Azruk, R. Nahoum, M. Guilloux-Viry, 
C. Thivet, A. Perrin, S. Labdi , Z. Z. Li and H. Raffy, 
Phys. Rev. B {\bf 49}, 9846 (1994). 
\bibitem{fivefriends} C. M. Varma, P. B. Littlewood, S. Schmitt-Rink, 
E. Abrahams and A. E. Ruckenstein, Phys. Rev. Lett., {\bf 63} 1996, (1989). 
\bibitem{ruckenstein}A. E. Ruckenstein and C. M. Varma, 
Physica C, {\bf 185-189}, 134, (1991). 
\bibitem{colemanioffetsvelik} 
P. Coleman, L. Ioffe and A. M. Tsvelik,  Phys. Rev. B {\bf 52}, 6611 (1995);
cond-mat/9504006.
\bibitem{threebody}P. Coleman, E. Miranda and A. M. Tsvelik, 
Phys. Rev. Lett. {\bf 74}, 1653 (1995); cond-mat/9406085. 
\bibitem{balatsky}E. Abrahams, A. Balatsky, D. J. Scalapino and 
J. R. Schrieffer, Phys. Rev. B, {\bf 52}, 1271 (1995); cond-mat/9503071. 
\bibitem{footy}Note that by choosing $\vec {\rm p}^*$ so that 
electrons and holes are degenerate to quadratic order, we will ensure 
that this Hamiltonian is diagonal to quadratic order. For the calculations 
we describe, an explicit form for the quadratic corrections to 
$\vec {\rm p}^*$ is not required. 
\bibitem{rammer86} J. Rammer and H. Smith, Rev. Mod. Phys. {\bf 58}, 
323 (1986). 
\bibitem{qp} One might argue that a scattering rate $\sim \omega$ is in 
contradiction to the assumption of a well defined quasi-particle peak 
in the spectral function. Under these circumstances one can integrate 
over momentum as is done for the high temperature electron-phonon 
systems to derive a similar transport equation. 
\bibitem{kadanoff} L. P. Kadanoff and G. Baym, {\it Quantum 
statistical mechanics}, W. A. Benjamin, New York (1962). 
\end{references}
\end{document}